%% file: cluster_CGM_ICM.tex
\documentclass[usenatbib]{mnras}
\usepackage{graphicx}
\usepackage{rotating}
\usepackage{pdfpages}
\usepackage{pdflscape}
\usepackage{longtable}
\usepackage{epstopdf}
\usepackage{threeparttable}
\usepackage{caption}
\usepackage{amsmath}

\newcommand{\mhaloeq}{M_{\rm halo}}

\newcommand{\msun}{$M_{\odot}$}
\newcommand{\msuneq}{M_{\odot}}
\newcommand{\rvir}{$r_{vir}$}

\newcommand{\hone}{H~\textsc{i}}

\newcommand{\ctwo}{C~\textsc{ii}}

\newcommand{\osix}{{\rm O}~\textsc{vi}}
\newcommand{\lya}{Ly$\alpha$}
\newcommand{\lyb}{Ly$\beta$}
\newcommand{\lyg}{Ly$\gamma$}
\newcommand{\lyd}{Ly$\delta$}

\newcommand{\limEW}{W$_{\rm lim}$}

\newcommand{\fc}{$f_c$}

\newcommand{\rtwo}{$r_{200}$}

\newcommand{\beq}{\begin{equation}}
\newcommand{\eeq}{\end{equation}}
\newcommand{\lam}{$\lambda$}
\newcommand{\lamnot}{$\lambda_O$}

\newcommand{\rband}{$r$-band}
\newcommand{\cmt}{{\rm cm}^{-2}}
\newcommand{\gmb}{GMBCG J255.3+64.23}
\newcommand{\maxb}{MaxBCG J217.8+24.68}

\def\gtrsim{\lower 2pt \hbox{$\, \buildrel {\scriptstyle >}\over
{\scriptstyle \sim}\,$}}
\def\lesssim{\lower 2pt \hbox{$\, \buildrel {\scriptstyle <}\over
{\scriptstyle \sim}\,$}}

\title[Warm-hot gas and depleted CGM in galaxy clusters]{Warm-hot gas in X-ray bright galaxy clusters and the H~I-deficient circumgalactic medium in dense environments}
\author[Joseph N. Burchett et al.]{Joseph N. Burchett$^{1}$$^{\dagger}$, Todd M. Tripp$^{1}$, Q. Daniel Wang$^{1}$, Christopher N.A. Willmer$^{2}$, \newauthor
David V. Bowen$^{3}$, Edward B. Jenkins$^{3}$ \\
$^{1}$ University of Massachusetts - Amherst \\ 
$^{2}$ Steward Observatory, University of Arizona \\
$^{3}$ Princeton University Observatory \\
$^{\dagger}$ Corresponding author: jburchet@astro.umass.edu}\pubyear{2017}

\begin{document}
\label{firstpage}
\maketitle

\begin{abstract}
We analyze the intracluster medium (ICM) and circumgalactic medium (CGM) in 7 X-ray detected galaxy clusters using spectra of background QSOs (HST-COS/STIS), optical spectroscopy of the cluster galaxies (MMT/Hectospec and SDSS), and X-ray imaging/spectroscopy (XMM-Newton and Chandra).  First, we report a very low covering fraction of \hone\ absorption in the CGM of these cluster galaxies, $f_c = 25^{+25}_{-15}\%$, to stringent detection limits (N(\hone) $<10^{13}~\cmt$). As field galaxies have an \hone\ covering fraction of $\sim 100\%$ at similar radii, the dearth of CGM \hone\ in our data indicates that the cluster environment has effectively stripped or overionized the gaseous halos of these cluster galaxies. Second, we assess the contribution of warm-hot ($10^5 - 10^6$ K) gas to the ICM as traced by \osix\ and broad \lya\ (BLA) absorption.  Despite the high signal-to-noise of our data, we do not detect \osix\ in any cluster, and we only detect BLA features in the QSO spectrum probing one cluster. We estimate that the total column density of warm-hot gas along this line of sight totals to $\sim3 \%$ of that contained in the hot $T>10^7$ K X-ray emitting phase.  Residing at high relative velocities, these features may trace pre-shocked material outside the cluster.  Comparing gaseous galaxy halos from the low-density `field' to galaxy groups and high-density clusters, we find that the CGM is progressively depleted of \hone\ with increasing environmental density, and the CGM is most severely transformed in galaxy clusters.  This CGM transformation may play a key role in environmental galaxy quenching.

\textbf{Keywords:}  galaxies: clusters: intracluster medium, galaxies: evolution, galaxies: haloes, quasars: absorption lines, galaxies: groups, X-rays: galaxies: clusters

\end{abstract}	

\section{Introduction}

Galaxy clusters represent the largest collapsed structures in the Universe.  In the $\Lambda$ Cold Dark Matter paradigm, these regions occur when strong density perturbations in the early Universe grow through structure formation as dark matter and baryonic matter accrete via gravitational collapse.  As a result, we observe galaxy clusters in the present epoch to be rather extreme environments, comprising highly evolved galaxy populations, gravitationally heated gas with $T \gtrsim 10^7$ K, and prodigious galaxy-galaxy mergers and interactions.  Therefore, clusters represent vital laboratories for studying how the largest structures are still forming from the cosmic web, how galaxy evolution is impacted by environment, and for testing the consistency of cosmological models and their measured parameters.  

Since the early cataloging of galaxy clusters \citep{Abell:1965fk}, the galaxy populations they host have been shown to possess quite different properties from galaxies in less dense environments.  Cluster galaxies are more likely to have quenched star formation, have elliptical morphologies, and be deficient in cold gas \citep[e.g.][]{Davies:1973lr,Oemler:1974fk, Dressler:1980qy,Sandage:1985jk}.  Within clusters, the distributions of galaxies with varying morphologies, star formation rate (SFR), and gas content have motivated much investigation into the mechanisms that transform galaxies.  To explain the proclivity of S0 and elliptical galaxies to reside cluster centres, \citet{Spitzer:1951lr} concluded that galaxy-galaxy collisions could remove the interstellar gas from normal spiral galaxies to leave behind S0s.  \citet{Gunn:1972qy} found that the cluster galaxies' gas reservoirs, out of which stars could form, should be removed by interactions with the hot intracluster medium (ICM) itself by ram pressure stripping.  Indeed, both galaxy-galaxy and galaxy-ICM interactions \citep[e.g.,][]{Wang:2004lr, Lu:2011fk,Roediger:2015qy} have been shown to be important transformation mechanisms in modern hydrodynamical simulations.  As an example of the former, \citet{Moore:1996ys} have argued that the cumulative effect of many high-speed galaxy-galaxy encounters can profoundly alter the gas distributions and morphologies of galaxies in the cluster environment, a process they refer to as `galaxy harassment'  \citep{Moore:1996ys,Moore:1998lr, Marasco:2016aa}.  Ram pressure stripping has been shown theoretically to be effective at removing a galaxy's halo gas well beyond the cluster virial radius \citep[e.g.,][]{Bahe:2013yu,Zinger:2016yq} as well as gas in the disk once the galaxy falls within the cluster virial radius \citep{Tonnesen:2007yq}.   The shock-heated ICM may also quench galaxies by halting the accretion of fresh fuel for star formation \citep[`strangulation';][]{Larson:1980vn,Voort:2017lr}.  Despite all of this work, the effectiveness of these processes has yet to be constrained observationally.

Observationally, 21 cm emission mapping of neutral hydrogen (\hone) reveals the evidence of ram pressure stripping, harassment, and tidal stripping due to the cluster itself acting on galaxies' disk gas \citep{Warmels:1988fk,Chung:2009fk}.  However, the effects of the cluster environment on the circumgalactic medium (CGM) -- massive gaseous reservoirs enveloping galaxies -- are considerably more difficult to observe due to the low densities and likely  temperatures of the CGM.  Currently absorption-line spectroscopy using background quasi-stellar objects (QSOs) provides the most practical means to investigate how the cluster environment affects the low-density CGM.  Most resonance transitions must be observed in the ultraviolet (UV) at the lower redshifts of mature galaxy clusters, and the relative scarcity of UV-bright background sources that are well positioned to probe foreground galaxies in clusters has resulted in limited data appropriate for such investigation.  In contrast, the \hone\ in the CGM of field galaxies has been characterized for two decades \citep[e.g.,][]{Lanzetta:1995rt, Tripp:1998kq, Chen:2005kx, Wakker:2009fr, Prochaska:2011yq, Tumlinson:2013cr, Burchett:2015aa}.  Preliminary results have recently suggested that the properties of circumgalactic \hone\ in denser environments differ significantly from the circumgalactic \hone\ of field galaxies. \citet{Yoon:2013kq} show that the CGM in their sample of Virgo cluster galaxies exhibit a decreased covering fraction of \hone\ relative to galaxies in the field.  Moreover, \citet{Burchett:2015aa} have found that environmental effects on the CGM may be evident even on group scales.

While the cooler, low-density phases of the CGM are most easily probed in absorption with current facilities, the hot $>10^6$ K phase of the ICM is readily observed in X-rays via bremsstrahlung and line emission, particularly in the central cluster regions.  In addition to the galaxy quenching/transformation phenomena discussed above, the X-ray emitting ICM provides key insight into other major astrophysical questions.  Among these is the so-called ``missing baryons" problem \citep{Persic:1992lr,Fukugita:1998vn}, whereby an accounting of the observed baryon mass relative to the dark matter mass at present times does not match the cosmological baryon fraction measured via, e.g., Big Bang nucleosynthesis as constrained by deuterium abundances \citep{Riemer-Sorensen:2017oj,OMeara:2006uk} or fluctuations in the cosmic microwave background \citep{Planck-Collaboration:2016eq}.  Indeed, the baryon fractions in galaxy clusters, where the baryons should be partitioned   primarily between the stars and X-ray emitting gas, have been shown to be deficient relative to the universal fraction \citep{Vikhlinin:2006zl,Andreon:2010fk,Lagana:2013qf}.  However, \citet{Gonzalez:2013lr}, identifying a mass-dependent contribution of the intracluster light, argue that the baryon content indeed appears to be closed for clusters $M_{500} > 3 \times 10^{14}~\msuneq$, but much scatter exists below these masses\footnote{Throughout this paper, $r_{500}$ refers to the radius within which the average mass density is 500 times the critical density of the Universe. M$_{500}$ is the mass enclosed within $r_{500}$.  Similarly, M$_{200}$ and \rtwo\ correspond to overdensities by a factor of 200.}.  Also, towards even lower masses (M$_{\rm halo} = 10^{13-14}M_{\odot}$), dynamically relaxed groups may have baryon fractions calculated from simply stars and hot gas consistent with the universal value \citep{Mathews:2005ys,Buote:2016qy}, although groups of similar masses may not exist in the same dynamical state and/or emit in X-rays at detectable surface brightness.   In any case, such conclusions only apply to the inner regions (e.g., $< r_{500}$).  Inference of the baryon content at larger radii, i.e., at $>r_{200}$, from X-ray observations depends sensitively on such assumptions as the clumpiness and thermal state of the hot gas. 

Cosmological hydrodynamical simulations suggest that the missing baryon contribution should arise from `warm-hot' gas at $T \sim 10^{5-6}$ K \citep{Cen:1999yq}, which would elude detection by most observational techniques.  The primary means by which observers can search for this warm-hot material with current facilities involve absorption line spectroscopy in the UV and X-ray regimes.  Detections of the warm-hot intergalactic medium\footnote{The term `WHIM' refers to intergalactic material residing in large-scale structures such as filaments and down to very low overdensities \citep[$\delta<50$;][]{Dave:2001dp}; we are focused on the regions within galaxy clusters and on the outskirts of galaxy clusters and will simply refer to gas with temperatures $T \sim 10^{5-6}$ K as `warm-hot'.} have been reported using X-ray absorption \citep[e.g.,][]{Fang:2002aa,Nicastro:2005aa,Buote:2009aa,Bonamente:2016aa}; however, many of these marginal detections have been controversial owing to the lack of high spectral resolution and sensitivity combined with systematic issues in X-ray spectra.  Gas in the $T \sim 10^{5-6}$ K regime has been readily detected with less ambiguity in UV absorption lines, namely the \osix\ $\lambda \lambda 1032,1038$ \AA\ doublet \citep{Tripp:2000aa,Danforth:2005aa,Tripp:2008lr, Savage:2014ty} and broadened \hone\ \lya\ features \citep{Tripp:2001fk,Richter:2006aa, Lehner:2007lr,Tejos:2016qv}. All of these studies have provided evidence that \osix\ and broad Lya absorbers could be important baryon reservoirs, but the physical conditions and origins of these absorption systems remain open topics of debate.  Likewise, UV absorption line spectroscopy can be employed to measure the warm-hot gas contribution to the baryon budget on galaxy cluster scales provided the sightlines are appropriately positioned. 

In this paper, we present analyses combining optical galaxy spectroscopy, UV QSO spectroscopy, and X-ray imaging and spectroscopy of five galaxy cluster systems (two of which are merging cluster pairs) to characterize the circumgalactic medium in the cluster environment as well as quantify the contribution of $T = 10^{5-6}$ K gas to the baryon content in the outer regions of galaxy clusters.  In Section \ref{sec:data}, we describe our observations and data.  Section \ref{sec:galaxies} presents the cluster galaxy data and analysis, and Section \ref{sec:results} contains the results gleaned from the full dataset.  Section \ref{sec:discussion} discusses the implications of our results, and we summarize our findings and conclude the paper in Section \ref{sec:summary}.

\section{Data and Observations}
\label{sec:data}

\subsection{Sample Selection and X-ray imaging/spectroscopy}
The key feature of our study is that we jointly probe a sample of galaxy clusters with X-rays and QSO absorption line spectroscopy.  Our core galaxy cluster/QSO sample was selected such that UV-bright background QSOs fall within 1.5 \rtwo\ of optically-selected, rich clusters at $z_c > 0.10$.  This redshift lower limit was imposed to place the \osix\ \lam 1032, 1038 doublet at the cluster redshift within the bandpass of spectrographs aboard the Hubble Space Telescope (HST).  

Two of these clusters (Abell 1095 and Abell 1926), were targeted in our HST/XMM-Newton joint program, HST GO 13342 (PI: Wang). The two QSO/cluster pairs observed in this program were selected such that the diffuse outer regions of clusters would be bright enough to observe with $XMM$-$Newton$ efficiently and that the QSOs had sufficient UV brightness to be observed with the Cosmic Origins Spectrograph (COS) in a reasonable number of HST orbits.  The third targeted cluster field, Abell 2246, was observed with the Chandra X-ray Telescope, and its observations were compiled from the Chandra archive. The A1095 and A1926 fields were observed with the $XMM-Newton$ European Photon Imaging Cameras (EPIC) for 47 ks and 47.5 ks, respectively.  The A2246 field was observed with the $Chandra$ X-ray telescope Advanced CCD Imaging Spectrometer (ACIS) for 240 ks.   Further details regarding the program design/target selection as well as the X-ray imaging and spectroscopy observations, data reduction, and measurements are provided by \citet{Wang:2014lr} and \citet{Ge:2016lr} (hereafter WW14 and G16, respectively).

\begin{figure*}
\includegraphics[width=0.99\linewidth]{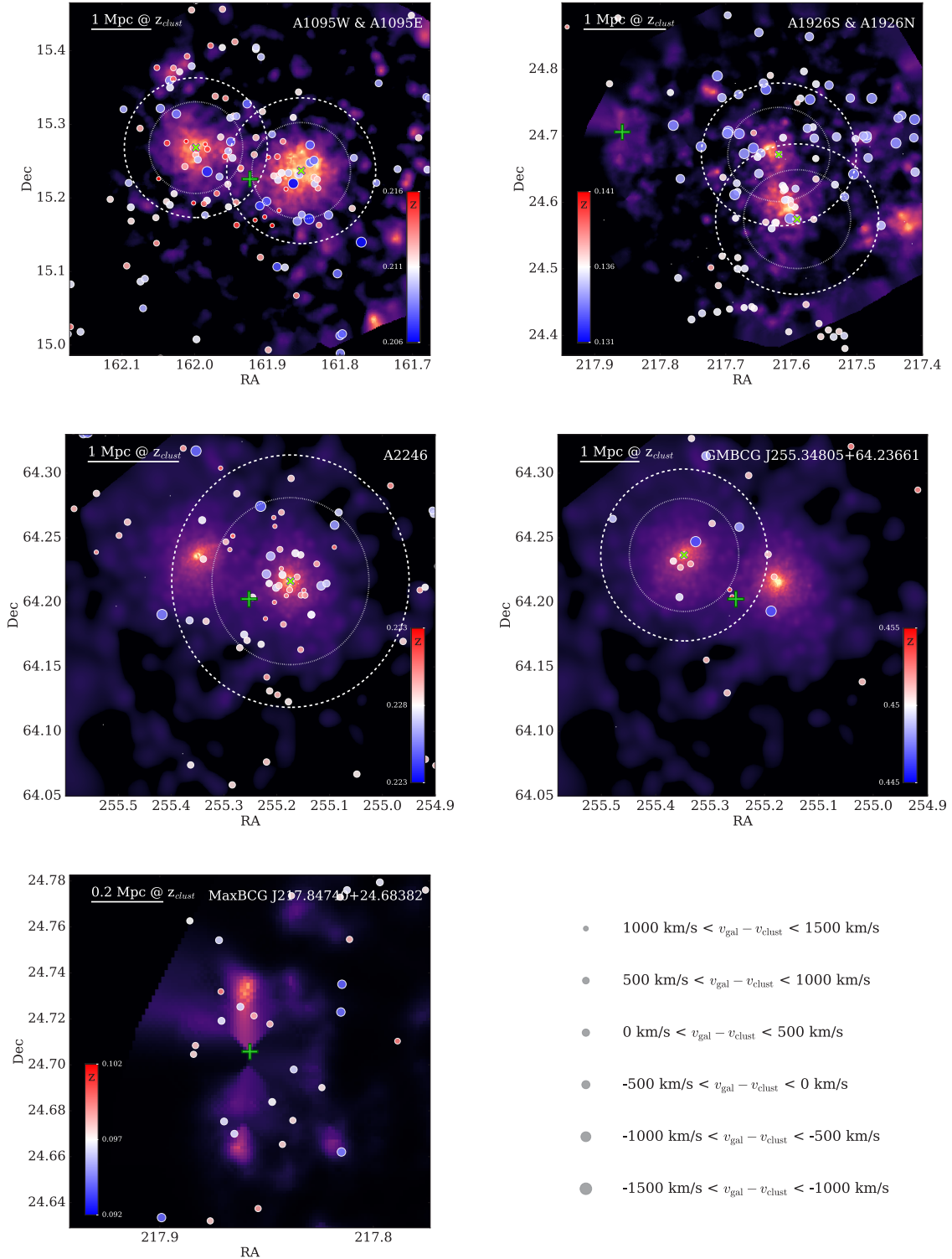}
\vspace*{-15mm}
\caption{Spectroscopically measured galaxies (filled circles) with redshifts within $\delta z = 0.01$ of the cluster brightest cluster galaxies plotted on sky.  The X-ray intensity is plotted in colour and is scaled to maximize visibility of low surface brightness emission; the X-ray centroids of each subcluster are marked with green crosses.  The large dashed and dotted circles mark \rtwo\ and $r_{500}$, respectively, derived from each X-ray derived cluster/subcluster centroid (Lying at the edge of the FOV, the X-ray data for \maxb\ are insufficient to derive masses, etc., so the virial radii are not marked for this cluster.).  The green plus signs denote the location of each QSO sightline (see Table \ref{tab:clusters}), around which our galaxy surveys were centered.  Corresponding to the colourbar in each panel, the filled circles are coloured according to galaxy redshift relative to each cluster redshift (marked at the centre of each colourbar).  Note: Even though both clusters fall within the same FOV, A2246 and \gmb\ as well as A1926N/S and \maxb\ are plotted separately due to their widely separated redshifts.}
\label{fig:XrayMapMontage}
\end{figure*}

The X-ray imaging for A1095 and A1926 (Figure \ref{fig:XrayMapMontage}) reveals rich substructure, most importantly, dual concentrations of emission suggesting that each `cluster' is actually a pair of merging subclusters.  G16 derive the physical properties (mass, temperature, etc.) of each individual subcluster, labeling them A1095W, A1095E, A1926S, and A1926N.  These X-ray measurements lead to substantially more reliable estimates of the cluster masses and hence their $r_{200}$ radii than the optical catalogs from which they were selected. They are significantly smaller than the original estimates based on the optical richnesses, largely due to the subcluster pairing. We adopt the same nomenclature when referring to these smaller clusters individually but simply refer to `A1095' and `A1926' when treating the merging system or targeted field as a whole.  In addition to the targeted clusters, other previously identified clusters from various optical catalogs were also detected in X-ray emission in our target fields.  We include two of these, GMBCG J255.34805+64.23661 (WW14) and MaxBCG J217.84740+24.68382 (G16), in our analysis and will refer to them as \gmb\ and \maxb, respectively.

Certain salient characteristics of the galaxy clusters featured in this study are summarized in Table \ref{tab:clusters}.  

\input{ClusterTable.tex}

\subsection{QSO spectroscopy}
The three QSOs included in this study were observed with HST using COS \citep{Green:2012qy} and STIS \citep{Woodgate:1998fj}.  We obtained COS observations of SDSS J104741.75+151332.2 and 2MASS J1431258+244220 through the joint XMM-Newton/HST program described above.  Additionally, SDSS J104741.75+151332.2 was observed with the G160M grating as part of HST GO program 13833 (PI: Tejos), and 2MASS J1431258+244220 was observed in program 12603. HS1700+6416 was observed with the COS G130M grating for the HST program 13491 (PI: Tripp) and with the STIS E140M grating for program 7778 (PI: Jenkins).  

\input{QSO_Obs_edit.tex}

COS far-ultraviolet spectra are recorded with two abutted cross-delay line microchannel-plate detectors \citep{CosHandbook2017}, and the physical gap between the two segments leads to an appreciable gap ($\approx$15 \AA) in spectral coverage. To fill in this gap in our programs, we observed the targets with two or three different grating tilts (for program 13342, we used central wavelengths = 1309 and 1327 \AA ; for program 13491 we employed CENWAVE = 1309, 1318, and 1327 \AA). In addition, to mitigate the effects of COS fixed-pattern noise, we obtained multiple exposures at different focal-plane split positions with each grating tilt.  As noted above, J104741.75+151332.2 was also observed for program 13833, which used CENWAVE = 1291 \AA\ for G130M and CENWAVEs = 1577 and 1600 \AA\ for G160M.  The G130M observations of J1431258+244220 for program 12603 used CENWAVEs = 1300 and 1327 \AA . The STIS observations of HS1700+6416 used the E140M echelle mode with the $0.2\times0.2$ aperture. To produce our final, fully combined spectra, we retrieved the data from the Mikulsky Archive for Space Telescopes and carried out the standard processing steps and 1-d spectrum extraction using CALCOS (v3.1.7) and CALSTIS (v2.22).  We then checked and (if necessary) adjusted the wavelength alignment of exposures by comparing appropriate absorption lines (e.g., Milky Way interstellar lines of comparable strength), and finally we coadded the spectra using the weighting scheme described by \citet{Tripp:2001fk} including coaddition of the overlapping regions of adjacent echelle orders in the STIS data. The fully coadded COS data cover wavelength ranges of 1133$-$1778 \AA\ (J104741.75+151332.2), 1150$-$1470 \AA\ (J1431258+244220) and 1152$-$1472 \AA\ (HS1700+6416) with spectral resolving power of 16000$-$21000 \citep{CosHandbook2017}.  The STIS E140M echelle spectrum of HS1700+6416 covers 1150$-$1710 \AA\ with a resolving power of 45800 \citep{StisHandbook2017}.  While the COS G130M and STIS echelle spectra of HS1700+6416 overlap at observed wavelengths $\lambda _{obs} <$ 1472 \AA , the signal-to-noise (S/N) of the COS G130M spectrum is much higher in this region, and we primarily use the STIS data for constraints on lines with $\lambda _{obs} >$ 1472 \AA .

The COS G130M spectra of SDSS J104741.75+151332.2 and 2MASS J1431258+244220 have S/N of approximately 24 and 14, respectively, per resolution element at the observed wavelength of the \osix\ doublet at the redshifts of our targeted clusters.  The G130M spectrum of 2MASS J1431258+244220 also covers \lya\ at S/N = 17.  The G130M spectrum of HS1700+6416 reaches a S/N of 40 near the expected wavelength of \osix\ arising from A2246.  The G160M spectrum of SDSS J104741.75+151332.2 and the STIS spectrum of HS1700+6416 are included here to cover \lya\ at the redshifts of A1095 and A2246, with S/N = 18 and 4, respectively.  The COS spectra and the STIS spectrum have spectral resolution $\sim$15 km s$^{-1}$ and $\sim$7 km s$^{-1}$, respectively.   A summary of our QSO observations and their corresponding archival datasets is presented in Table~\ref{tab:QsoObs}.

\subsection{Galaxy Spectroscopy}
\label{sec:galSpec}
Given the redshifts of our target QSO-cluster pairs (WW14 \& G16), we required a deeper spectroscopic galaxy survey than the publicly available data in order to investigate CGM properties in the clusters.  Therefore, in addition to data from SDSS, we obtained new observations at the MMT with the Hectospec Multifiber Spectrograph \citep{Fabricant:2005fk,Mink:2007lr}.  Hectospec is a 300 fiber spectrometer with a 1$^\circ$ diameter field of view (FOV) and covers a wavelength range of 3650-9200 \AA\ with a 6 \AA\ resolution (1.2 \AA~pixel$^{-1}$, R = $600-1500$).  

We selected galaxies in the cluster/QSO fields to receive fibers using SDSS photometry according to the following procedure.  First, we queried the SDSS for all objects identified as galaxies within 30 arcmin of each QSO sightline and that would have \rband\ absolute magnitudes corresponding to at least 0.1 $L*$ \citep[$\mathcal{M}_r \lesssim -18.6$;][]{Blanton:2003kx} if at the redshift-indicated distances of the targeted clusters.  We then removed objects with photometrically measured properties that suggested problematic SDSS data.  These included making a cut on the \rband\ `\mbox{fiberMag}' ($r_{\rm fib} \leq 22.8$) to remove objects with magnitudes far fainter than are reliable for SDSS photometry; such faint magnitudes are symptomatic of image artifacts being classified as galaxies.  Also, we found that image artifacts and other clearly misclassified or problematically measured objects in the SDSS also contain unusually large \rband\ Petrosian radius measurements (`\mbox{petroRad\_r}' $>$ 100 arcmin).  Lastly, following \citet{Strauss:2002fk}, we rejected objects with \rband\ `\mbox{psfMag}' values within 0.1 mag of their `modelMag' values to mitigate confused star/galaxy classifications.  To further reject objects that were likely stars or extremely distant galaxies ($z_{\rm gal} >> z_{\rm QSO}$), we placed a conservative $z_{\rm QSO}$-dependent cut in $g$-$r$ --- $r$-$i$ colour-colour space based on archival spectroscopic galaxy surveys  \citep{Geha:2017aa}. We placed colour thresholds such that galaxies in these surveys with $g$-$r$ and $r$-$i$ colours redder than the threshold values were likely to have  $z_{\rm gal} > z_{\rm QSO}$.  For fiber targeting, we then selected all remaining galaxies without SDSS spectra and with \rband\ absolute magnitudes corresponding to 0.2 $L*$ or brighter if at the redshifts of the targeted clusters. 

\begin{table}
\begin{center}
\footnotesize
\caption{Summary of MMT/Hectospec Observations}
\vskip 0.1cm
\begin{tabular}{lrrrl}
\hline
\hline
Field  & R.A. (J2000)      & Dec. (J2000)  & Exposure  & Date\\
\hline
A1095  & 10:47:41.33 & +15:13:44.30 & 5 x 780s & 2016-02-08\\
A1095  & 10:47:41.33 & +15:13:44.30 & 4 x 720s & 2016-03-12\\
A1926  & 14:31:21.92 & +24:42:02.68 & 5 x 600s & 2016-02-04\\
A1926  & 14:31:21.92 & +24:42:02.68 & 4 x 600s & 2016-03-15\\
A2246  & 17:00:58.80 & +64:12:55.33 & 5 x 600s & 2014-06-24\\
A2246  & 17:00:57.31 & +64:13:09.67 & 5 x 900s & 2014-06-25\\
A2246  & 17:01:00.27 & +64:13:14.49 & 5 x 720s & 2014-06-25\\
A2246  & 17:01:05.22 & +64:12:57.12 & 5 x 960s & 2016-03-02\\
A2246  & 17:01:05.22 & +64:12:57.12 & 5 x 960s & 2016-03-03\\
\hline
\label{tab:GalaxyObs}
\end{tabular}
\end{center}
\end{table}

The assignment of objects to fibers was done using XFITFIBS {\footnote{\url{http://www.harvard.edu/john/xfitfibs}}, which takes into account the number of configurations, the object priorities, and the number of sky positions. For the observations here, we assigned a minimum of 40 fibers to measure the sky background and 10 to place on stars for flux calibration \citep{Cool:2008qy}, leaving typically 250 fibers that could be assigned to program objects.  XFITFIBS enables the survey designer to assign priorities for objects to receive fibers in the generated configurations, and we prioritized galaxies based on their QSO-centric impact parameters ($\rho$), impact parameters relative to the cluster X-ray centroids measured by WW14 and G16, and colour consistency with belonging to a `red sequence' \citep[e.g.,][]{Faber:1973aa} with other cluster galaxies.  The colour criteria for red sequence candidates were set according to visually identified overdensities in $g-r$ colour for each cluster field. We assigned priorities for each targeted galaxy in the following order: 1) impact parameters $\rho < 300$ kpc from the QSO sightline, 2) red sequence candidates with cluster-centric $\rho < 3$ Mpc, 3) non-red sequence candidates with cluster-centric $\rho < 3$ Mpc, 4) red sequence candidates and cluster-centric $\rho > 3$ Mpc, and 5) non-red sequence candidates with cluster-centric $\rho > 3$ Mpc.

The Hectospec observations were taken in queue mode, which enables targeting the field when optimally placed on the sky. A total of 9 configurations were obtained, 3 during Spring 2014 and 6 during Spring 2016. A log of the observations is shown in Table \ref{tab:GalaxyObs}, which contains the field identifications (named for the targeted cluster), J2000.0 coordinates, exposure times, and dates of observation.  

The Hectospec data were reduced using the {\it{HSRED 2.0}} pipeline \citep{Cool:2008qy}, which is an ensemble of $IDL$ scripts based on the SDSS pipeline. {\it{HSRED}} performs bias, flatfield, illumination, and wavelength calibrations, subtracts the sky background, and extracts one-dimensional spectra.  The flux calibration was done using the spectra of 6Ð10 stars selected to have SDSS colours consistent with F stars and that are observed simultaneously with the main galaxy sample. The flux calibration correction was obtained combining the extinction-corrected SDSS photometry of these stars with \citet{Kurucz:1993uq} model fits \citep{Cool:2008qy}. These stellar spectra were also used to remove the telluric lines. The spectral range covered by Hectospec allows for the detection of one or more typical emission lines present in the spectra of galaxies ([O~\textsc{ii}], H$\beta$, [O~\textsc{iii}], H$\alpha$), for galaxies to $z\sim1$. The redshifts measured by HSRED also use a code adapted from SDSS and the same templates as SDSS \citep{Stoughton:2002ec}. All spectra were visually inspected for validation as described below.  

A redshift quality flag is assigned to each spectrum, following the same procedure used for the DEEP2 survey \citep{Willmer:2006lr,Newman:2013fj}, where redshift qualities range from Q = 4 (probability P $>$95\% of being correct), 3 (90\% $<$ P $<$95\%), 2 (P $<$90\%), and 1 (no features recognized). Q = 2 spectra are assigned to objects for which only a single feature is detected, but cannot be identified without ambiguity. The Q = 3 spectra have more than one spectral feature identified, but tend to have low S/N. The typical confidence levels for these objects is ~90\% for the DEEP2 galaxies. Finally, Q = 4 objects have two or more spectral features with reasonable to high S/N. The confidence level of these redshifts is typically $>$95\%. Because of the larger spectral range covered by Hectospec (3800-9500 \AA) relative to DEEP2 (5000-9500 \AA), we expect that the quoted confidence levels are the conservative limits for our spectra.  HSRED returns typical redshift measurement errors of $\sigma_{z} \sim 0.0001$. However, upon evaluating the dispersion in redshift measurements for several objects with duplicated Hectospec observations, we adopt a more conservative $\sigma_{z} \sim 0.00055$ ($\sigma_v \sim165$ km s$^{-1}$) as our minimum uncertainty.

For the A2246 field, we also include galaxies measured using LDSS-2 on the William Herschel Telescope at the Observatorio del Roque de Los Muchachos.  Details regarding the observations and redshift measurement techniques are provided by \citet{Bowen:2001rf}.  

In the Appendix, we provide all galaxy redshifts obtained through our spectroscopic efforts as well as those included in our analysis from SDSS.

\section{Cluster Member Galaxies}
\label{sec:galaxies}
The optical galaxy spectroscopy obtained in these QSO/fields dramatically expands the breadth and depth of analyses enabled by the X-ray and UV data.  The advantages of spectroscopically measured redshifts are many for studies of this kind.  First, photometric redshifts provided by the SDSS have typical uncertainties on the order of $z\pm0.01$, corresponding to velocity uncertainties of thousands of km s$^{-1}$ and cosmological distance uncertainties of $\sim$50 Mpc at $z=0.2$.  In contrast, our spectroscopic redshifts have uncertainties $\delta z \sim 0.0001$, resulting in galaxy velocity uncertainties similar to those of the UV QSO spectra and enabling localization in velocity between individual galaxies and absorption features.  \citet{Tumlinson:2013cr} showed that \hone\ absorption systems associated with $\sim L*$ galaxy halos are typically found within $\pm$ 200 km s$^{-1}$ of the galaxy systemic velocities, setting an expected galaxy-absorber velocity separation scale.  We note that this velocity scale is smaller than typical cluster velocity dispersions by a factor of a few, and broadband photometric redshifts do not provide sufficient precision.  Second, the spectroscopic redshift precision mitigates confusion between legitimate cluster members and distant galaxies in the foreground and background.  Therefore, the velocity structure of the cluster can be more accurately characterized.  Although a full kinematic substructure analysis of these clusters is beyond the scope of the current paper, we include in Table \ref{tab:clusters} an estimate of each subcluster's velocity dispersion $\sigma_{cl}$.  We adopt the biweight scale estimator as described by \citep{Beers:1990lr} and calculate 1-$\sigma$ uncertainty estimates of approximately 50 km s$^{-1}$.

\begin{figure*}
\includegraphics[width=1.11\linewidth]{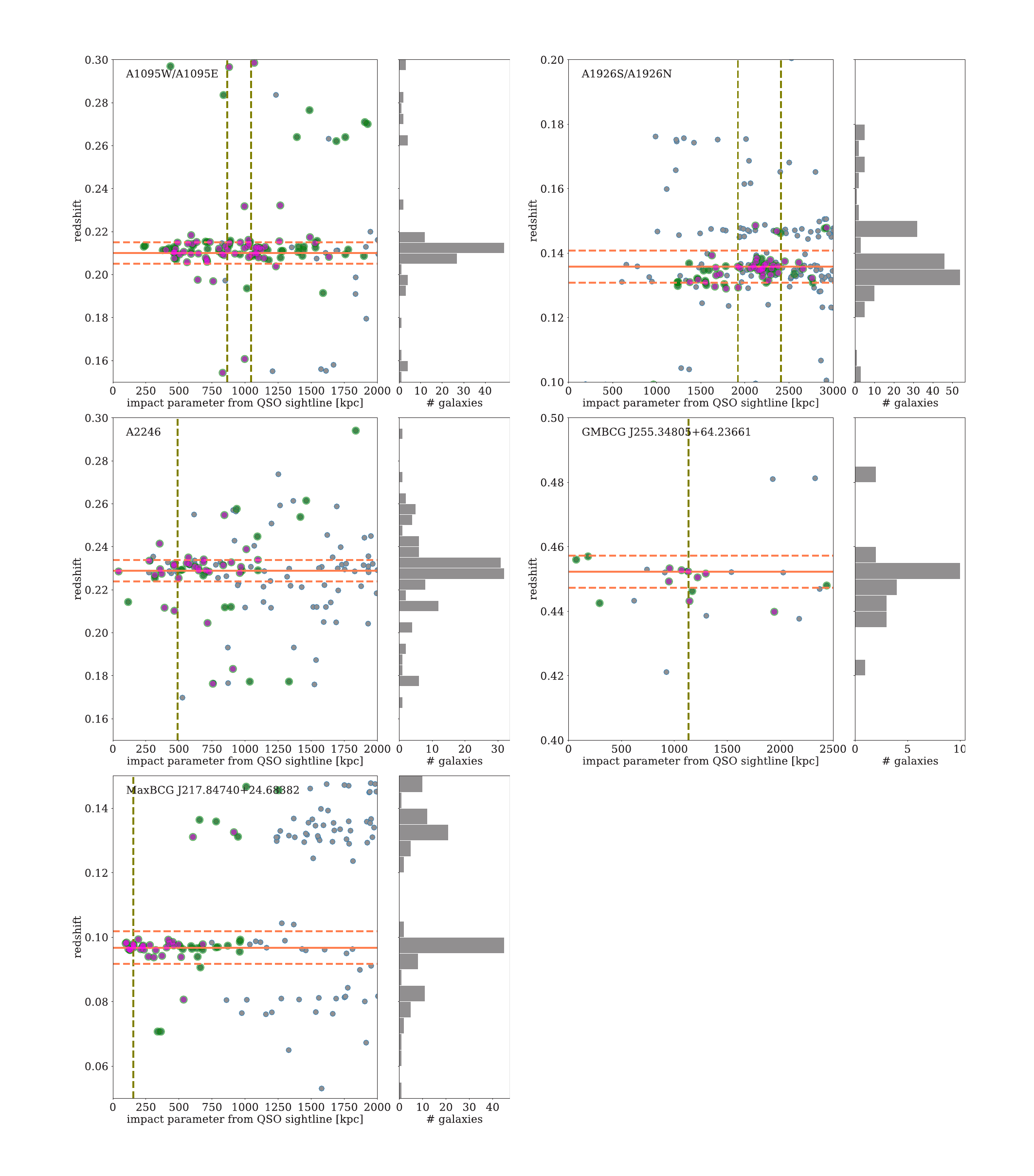}
\vspace*{-15mm}
\caption{Redshifts of spectroscopically measured galaxies as a function of impact parameter from the QSO sightlines probing each galaxy cluster.  The solid orange horizontal lines mark the cluster redshifts as reported by \citet{Wang:2014lr} and \citet{Ge:2016lr} and are derived from the optically identified BCGs; the only exception is A2246, for which our optical survey measured the redshift of the likely true BCG coinciding with the X-ray centroid.  Galaxies coloured in green and magenta fall within \rtwo\ and $r_{500}$, respectively, in projection of the cluster centroids.  The red dashed horizontal lines mark the redshift range of galaxies plotted in Figure \ref{fig:XrayMapMontage}.  The vertical dashed lines mark the projected separations of each X-ray centroid from the QSO sightline. To the right of each redshift map, histograms in bins of $\Delta z = 0.005$ show the redshift distributions of galaxies in each cluster region.  Our optical surveys in these regions have increased the number of galaxies with spectroscopic redshifts by more than a factor of 8 in each field.}
\label{fig:RadDistMontage}
\end{figure*}

Figure \ref{fig:RadDistMontage} shows the distributions of galaxies local to the targeted clusters.  WW14 and G16 measured X-ray centroids of these clusters (except for \maxb), and the cluster redshifts they provide correspond to spectroscopic redshifts for the brightest cluster galaxies (BCGs) identified in cluster catalogs from the literature.  Shown in Figure \ref{fig:RadDistMontage} are the spectroscopic redshifts from our survey plotted as a function of projected angular separation from the QSO sightlines probing our cluster sample.  The red solid horizontal lines mark the cluster redshifts given by WW14 and G16 and the red dashed lines mark the range in redshift of galaxies plotted in Figure \ref{fig:XrayMapMontage}.   Histograms of galaxy redshifts in bins of $\Delta z = 0.005$ are also given for each cluster field.   Our follow-up survey has dramatically increased the available spectroscopic measurements in these fields; by comparison, all of these clusters are contained within the SDSS footprint, but fewer than 1/8 of the spectra indicated in Figure \ref{fig:RadDistMontage} for any field are provided by SDSS.  The galaxy overdensities are quite conspicuous about the cluster redshifts and as shown by the histograms on the right of each panel, our survey reveals $>50$ cluster member candidates for most fields.   

\begin{figure}
\includegraphics[width=0.9\linewidth]{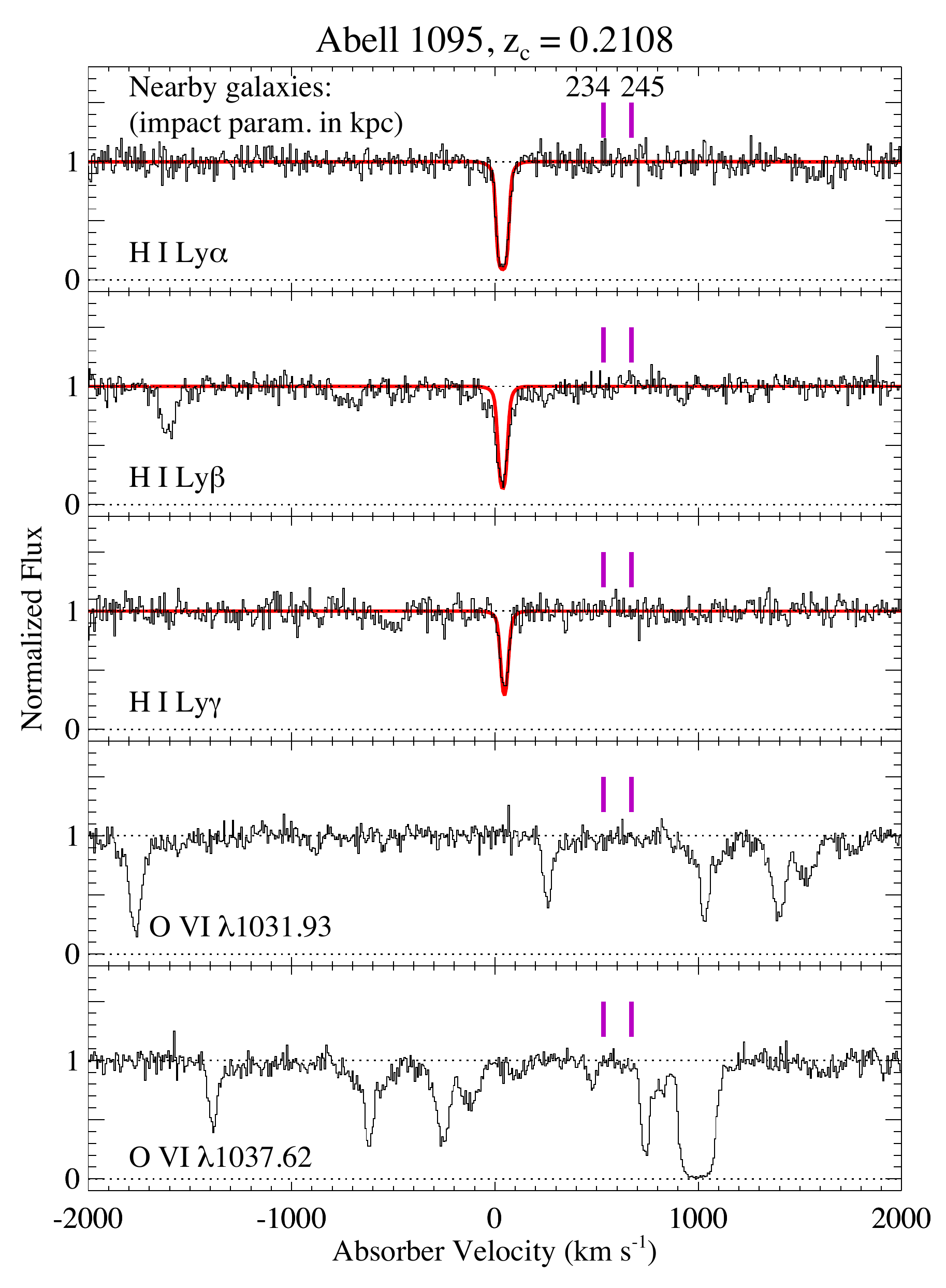}
\caption{The continuum-normalized COS spectrum of SDSS J104741.75+151332.2 showing the \hone\ and \osix\ transitions covered.  The velocity zero point is fixed to the optically derived redshift, $z_c$, of A1095 given by G16.  Voigt profiles fitted to the absorption lines identified within the velocity range shown are plotted in red.  Purple hashes mark the velocity offsets of galaxies within 300 kpc of the QSO sightline.} 
\label{fig:Stackplot1}
\end{figure}

\begin{figure}
\includegraphics[width=0.9\linewidth]{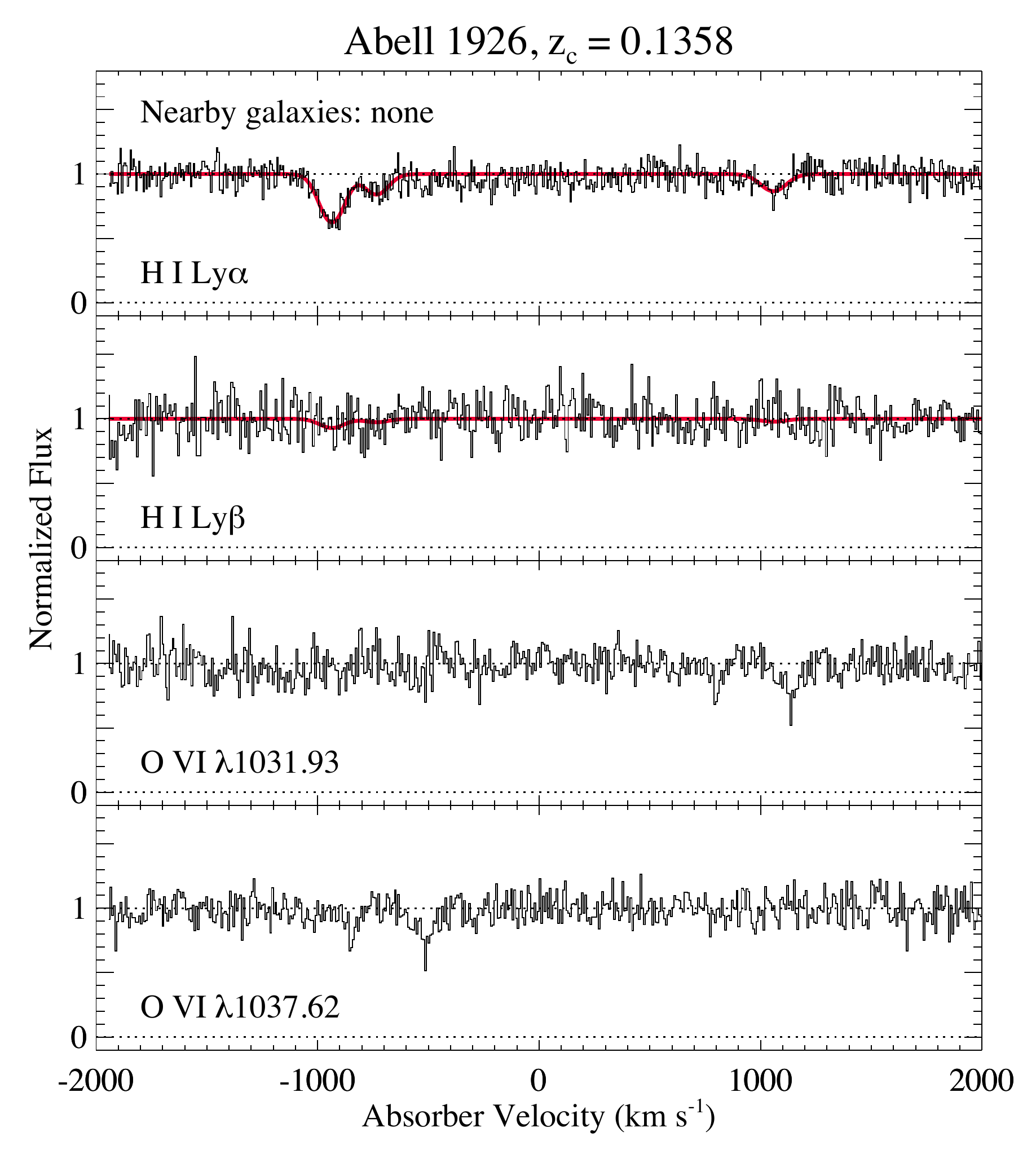}
\caption{The continuum-normalized COS spectrum of 2MASS J1431258+244220 showing the \hone\ and \osix\ transitions covered.  The velocity zero point is fixed to the optically derived redshift, $z_c$, of A1926 given by G16.  Voigt profiles fitted to the absorption lines identified within the velocity range shown are plotted in red.} 
\label{fig:Stackplot2}
\end{figure}

\begin{figure}
\includegraphics[width=0.9\linewidth]{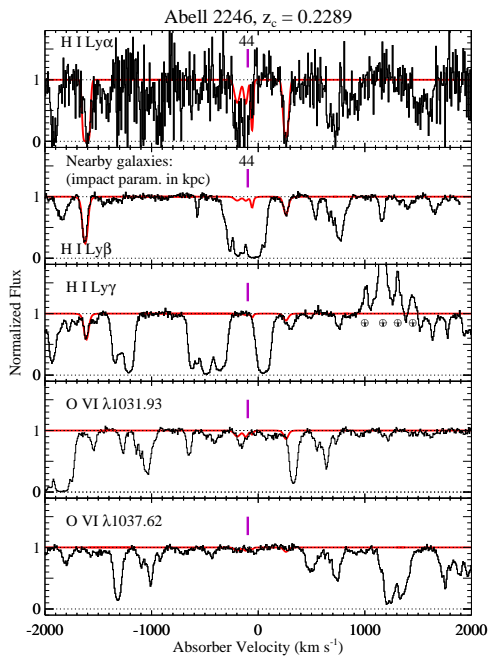}
\caption{The continuum-normalized COS and STIS spectra of HS1700+6416 showing the \hone\ and \osix\ transitions covered.  The velocity zero point is fixed to the optically derived redshift, $z_c$, of A2246 given by WW14.  Voigt profiles fitted to the absorption lines identified within the velocity range shown are plotted in red. Purple hashes mark the velocity offsets of galaxies within 300 kpc of the QSO sightline.  Note that the profiles shown between -200 and 100 km s$^{-1}$ were fitted to provide upper limits for the corresponding species; the \hone\ component at approx. +260 km s$^{-1}$ serves as the only unambiguous detection plotted here.  Geocoronal emission features in the centre panel are labeled with~$\oplus$.}
\label{fig:Stackplot3}
\end{figure}

\begin{figure}
\includegraphics[width=0.9\linewidth]{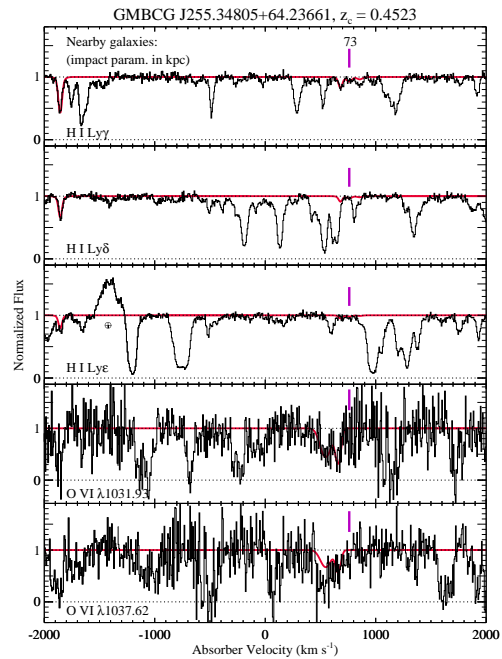}
\caption{The continuum-normalized COS and STIS spectra of HS1700+6416 showing the \hone\ and \osix\ transitions covered.  The velocity zero point is fixed to the optically derived redshift, $z_c$, of \gmb\ given by WW14.  Purple hashes mark the velocity offsets of galaxies within 300 kpc of the QSO sightline.  Note that the profiles shown in red were fitted to provide upper limits for the corresponding species only.  No absorption associated with this cluster was unambiguously detected.  A geocoronal emission feature in the centre panel is labeled with~$\oplus$.}
\label{fig:Stackplot4}
\end{figure}

\begin{figure}
\includegraphics[width=0.9\linewidth]{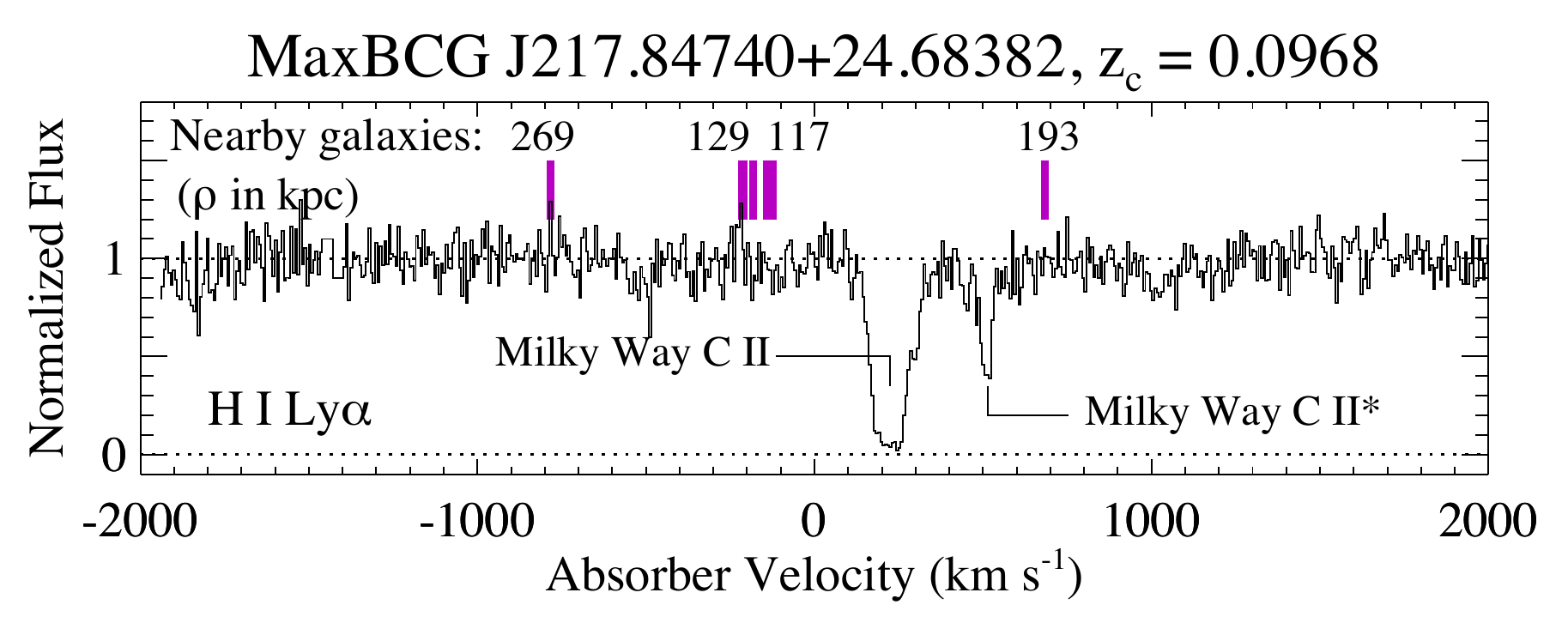}
\caption{The continuum-normalized COS and STIS spectra of 2MASS J1431258+244220 showing the \hone\ \lya\ transition.  The velocity zero point is fixed to the optically derived redshift, $z_c$, of \maxb\ given by G16.  Purple hashes mark the velocity offsets of galaxies within 300 kpc of the QSO sightline.}
\label{fig:Stackplot5}
\end{figure}

Figure \ref{fig:XrayMapMontage} shows the spectroscopically measured galaxies from our surveys and the SDSS within $\Delta z = 0.005$ of the BCG cluster redshifts along with the imaged X-ray emission.  The green cross in each panel denotes the position of the observed QSO sightline, and the dashed circles demarcate \rtwo\ from each cluster X-ray centroid as reported by WW14 and G16.  Note that A2246 and \gmb\ lie at quite different redshifts and are thus shown in separate panels though they fall within the same $Chandra$ FOV.  Likewise, A1926S/A1926N and \maxb\ are plotted separately due to their widely separated redshifts, and the colour scaling for the X-ray emission has been scaled separately to bring out low surface brightness features in each panel.

We defer to a subsequent publication several in-depth analyses involving the cluster member sample, including inferring the velocity and spatial substructure to compare with the X-ray morphology as well as analyzing the star formation activity in and around the clusters, and primarily focus on the UV absorption herein.

\section{Results}
\label{sec:results}

We now leverage the QSO absorption spectra along with the optical and X-ray data presented above to characterize the CGM in the cluster environment and investigate the warm-hot phase contribution to the ICM.  Figures \ref{fig:Stackplot1} - \ref{fig:Stackplot5} show continuum-normalized absorption profiles of \hone\ and \osix\ transitions covered in our QSO spectra centred about each cluster's redshift (corresponding to $v=0$).  In particular, the redshifts adopted here for $v=0$ are those spectroscopically measured for the BCG of the corresponding cluster or subcluster with the smallest impact parameter.  The red curves show Voigt profiles fitted to species that we report as either detections or upper limits.  For the latter cases, possible absorption arises from the species named but is heavily blended with interloping lines from other redshifts  (heavy blending precludes secure identification and measurement of \hone\ and \osix, so we conservatively derive upper limits on the maximum \hone/\osix\ absorption allowed by the data).   The purple ticks mark relative velocities of galaxies found at small impact parameters to the sightline (see the appendix for additional information on the galaxy-redshift survey).

\input{LineMeasurementTable_VP.tex}

In Table \ref{tab:ICMmeas}, we present the measurements for the \hone\ absorbers we detect and limits where absorption at the cluster redshift is blended with absorption lines from other redshifts.  For these possible blends, we fit Voigt profiles to all lines confidently identified in the relevant spectral region and then add components of \hone\ (using as many Lyman series lines as covered) until all of the optical depth was accounted for; in many cases, the absorption is predominantly due to lines from other redshifts, and by using the full set of available lines, we can set useful upper limits on \hone\ and \osix\ absorption despite the presence of these blends.  We report no unambiguous detections of \osix\ associated with any of the clusters.

\subsection{The CGM of Cluster Galaxies}
\label{sec:resultsCGM}
As shown in Figures \ref{fig:XrayMapMontage} and \ref{fig:RadDistMontage}, our galaxy surveys in the cluster fields reveal several galaxies at similar redshifts to the galaxy clusters and at small impact parameters.  Therefore, we exploit our QSO spectroscopy and galaxy survey data to produce a sample of galaxy/sightline pairs, for which we can measure the absorption at each galaxy redshift, enabling us to constrain the circumgalactic gas of galaxies residing in the cluster environment.

We begin by searching each sightline's spectroscopic galaxy database for objects within $\Delta v < 1200$ km s$^{-1}$ of the redshift of clusters probed.  Our redshift separation criterion balances selecting galaxies that are plausibly cluster members ($<$3$\sigma_{cl}$) and reducing the likelihood of selecting those separated by very large distances if the velocity separation were due to the Hubble flow.  This search produces an initial sample of several galaxies, whose velocities relative their host clusters are labelled in Figures \ref{fig:Stackplot1} - \ref{fig:Stackplot5} by magenta tick marks.  In most cases, the impact parameters of the galaxies are listed above the purple tick marks (except in the MaxBCG J217.84740+24.68382 plot, where many galaxies are close to the sightline; in this case, we only indicate the impact parameters of the closest galaxies). Then, we search for absorption associated with these galaxies in the HST/COS spectrum of the background QSO.  As we will compare our galaxy cluster CGM sample with the field sample of \citet{Prochaska:2011yq}, we adopt a similar galaxy/absorber velocity separation criterion: $|v_{\rm gal} - v_{\rm abs}| \leq 400$ km s$^{-1}$.   We note that the velocity window used by COS-Halos to draw galaxy-absorber associations is larger ($|\Delta v| \leq 600$ km s$^{-1}$); however, for all of their galaxies with associated absorption at $|\Delta v| \geq 400$ km s$^{-1}$, they also detect absorption components at $|\Delta v| \leq 400$ km s$^{-1}$.   Unfortunately, line blending with interloping absorption systems from different redshifts hampers stringently imposing these criteria.  We therefore detail the galaxy/absorber sample selection for each cluster individually below.  Our analysis will focus on galaxy/absorber associations with impact parameters $\rho < 300$ kpc; A1926 is the only cluster without spectroscopically confirmed galaxies within this impact parameter range.  We adopt $W / \sigma_{W} > 3$, where $W$ and $\sigma_W$ denote the equivalent width and its error, respectively, as our criterion for statistically significant absorption detections. Our CGM absorber measurements are collected in Table \ref{tab:CGMmeas}.  \\ \\
\emph{A1095:}  We detect galaxies with impact parameters 234 and 245 kpc.  No \hone\ or \osix\ absorption features arise in the QSO spectrum within 400 km s$^{-1}$ of these galaxies' redshifts (see Figure \ref{fig:Stackplot1}, although a narrow \hone\ component does lie at $\Delta v \sim 500$ km s$^{-1}$ from the galaxy at $\rho = 234$ kpc.  Due to the large velocity separation, we do not consider this absorber to be associated with either of the two galaxies, although we do offer an alternative explanation for its origin in Section \ref{sec:discussBLA}.  As no absorption was associated with these galaxies, we measured $3\sigma$ upper limits on the column density using the apparent optical depth method \citep[AODM;][]{Savage:1991vn} at their redshifts. \\ \\
\emph{A2246:}   A galaxy is detected with $\rho = 43$ kpc as is one unambiguous \hone\ component at $\delta v \sim 350$ km s$^{-1}$.  As the sightline probing A2246 and \gmb\ is that of a relatively high redshift QSO, many interloping systems appear in the spectral regions of interest. To place upper limits on \hone\ affiliated with the galaxy at $\rho = 43$ kpc, we used our Voigt-profile fitting software to determine the maximum amount of \hone\ absorption that is \emph{jointly} allowed by the \lya, \lyb, \lyg, \lyd, and Ly$\epsilon$ profiles. Comparison of these profiles revealed a clear detection of \hone\ at $v$ = 460 km s$^{-1}$ (indicated by fully consistent absorption lines of \lya\ and \lyb, which are not blended at this velocity, and corroborated by consistent optical depth in the higher Lyman series lines) with log N(\hone) = $13.97\pm0.02~\cmt$. The rest of the absorption within $\pm600$ km s$^{-1}$ of this galaxy is \emph{not} consistently present in the various Lyman series lines and therefore is at least partially due to absorption lines at other redshifts, but some affiliated \hone\ could be hidden in these nearby absorption blends. To determine how much \hone\ could be hidden in these blends, we iteratively explored models with one to several additional components, and we found that we obtained a maximal increase in \hone\ with three additional components that would contribute an additional \hone\ column density of log N(\hone) = $13.83~\cmt$.  We emphasize that due to blending, the three additional components cannot be securely attributed to \hone\ but are not ruled out by the data. Thus, we conservatively state that the \hone\ column is at least log N(\hone) = $13.97~\cmt$ and could be as high as log N(\hone) = $14.20~\cmt$.  For \osix, none of the absorption shows the velocity spacing and relative strengths of the \osix\ doublet. There are several features near the expected wavelength of the O VI 1031.92 line, but the O VI 1037.62 line falls in a relatively clean region.  Assuming that the \osix\ has a similar component structure to the \hone\ \citep[as is often observed in \osix\ absorbers,][]{Tripp:2008lr}, we derive log N(\osix) $<13.80~\cmt$ from the joint constraints provided by both lines of this doublet. \\  \\
\emph{\gmb :} We detect one galaxy at 73 kpc.  Like A2246, this cluster is probed by the HS1700+6416 sightline, and similar line blending issues arise. Unlike, A2246, no unambiguous \hone\ is identified near the redshift of \gmb.  Some very weak components may be present near the redshift of the detected galaxy, but these cannot be verified and we use a similar Voigt profile fitting procedure as that for A2246 above to place upper limits for \hone\ and \osix. \\ \\
\emph{\maxb :}  This cluster provides several galaxies within our impact parameter selection range but introduces complications due to line blending as well.  Unfortunately, \lya\ is the only \hone\ line covered in the COS bandpass due to the lower redshift of this cluster, and \osix\ is also not covered.  Also, the strong \ctwo\ and \ctwo$^*$ profiles from our own Galaxy fall very near the would-be location of \lya\ at the cluster redshift. The redshifts of several galaxies close to the sight line would place their \lya\ lines directly on top of the MW ISM  \ctwo/\ctwo$^*$ lines if the \lya\ lines are exactly at the galaxies' systemic velocities.  Since the Galactic \ctwo\ is saturated, it would be difficult to recognize the \lya\ lines if this were the case.  Of course, affiliated absorption does not necessarily occur exactly at a galaxy's systemic redshift, and we would detect \lya\ from these galaxies at other velocities, but to be conservative we omit these galaxies from the CGM/absorber sample CGM galaxy/absorber sample, keeping only those shown at $v_{\rm abs} < 0$ km s$^{-1}$ at $v_{\rm abs} > 600$ km s$^{-1}$ in Figure \ref{fig:Stackplot5}.  We have identified the weak absorption feature seen at $v_{\rm abs} \sim -600$ km s$^{-1}$ in Figure \ref{fig:Stackplot5} as Ly$\delta$ at $z=0.3686$ (the corresponding Ly$\beta$ is also covered and detected with a proportionately strong profile).  Although we concede that \lya\ absorption could arise at $v_{\rm abs} - v_{\rm gal} \leq 400$ km s$^{-1}$ from those galaxies we have kept in our sample, this absorption would be blended with the Milky Way lines and have likelier associations with those galaxies we have omitted.  Again, 3$\sigma$ upper limits are derived by AODM at the redshifts of each galaxy remaining in our sample.  One galaxy at $z\sim0.0963$ yielded $W / \sigma_{W} \sim 3.25$ when integrating over $v \pm 50$ km s$^{-1}$ about its redshift, and the \hone\ measurement for this galaxy in Table \ref{tab:CGMmeas} reflects the integrated apparent column density and its error. \\ \\

\input{LineMeasurementTable.tex}

Figure \ref{fig:HIprofile} shows the resulting N(\hone) as a function of QSO-galaxy impact parameter constructed from our cluster CGM sample and two surveys from the literature, COS-Halos \citep{Tumlinson:2013cr} and \citet{Prochaska:2011yq}.  The \hone\ absorption appears to be highly suppressed in the CGM of our cluster galaxies relative to the literature data.  For the eleven galaxies shown here, we detect \hone\ absorption with $\rho<$ 300 kpc  of only two.  For our detection limits \limEW $> 30$ m\AA, we report a covering fraction of $f$(\hone) = $25^{+25}_{-15}\%$ at $\rho<$ 300 kpc of our cluster galaxies.  We quantitatively compared our data with that of COS-Halos using a Gehan's Generalized Wilcoxon Test to calculate that probability that the column densities we measure for the sightlines overlapping in impact parameter with COS-Halos ($<160$ kpc) could have been drawn from the same column density distribution.  According to this statistical test, which is well-suited to accommodate censored data, we may reject the null hypothesis with 99.7\% confidence that the two samples were drawn from the same parent population.

\begin{figure} 
\includegraphics[width=1\linewidth]{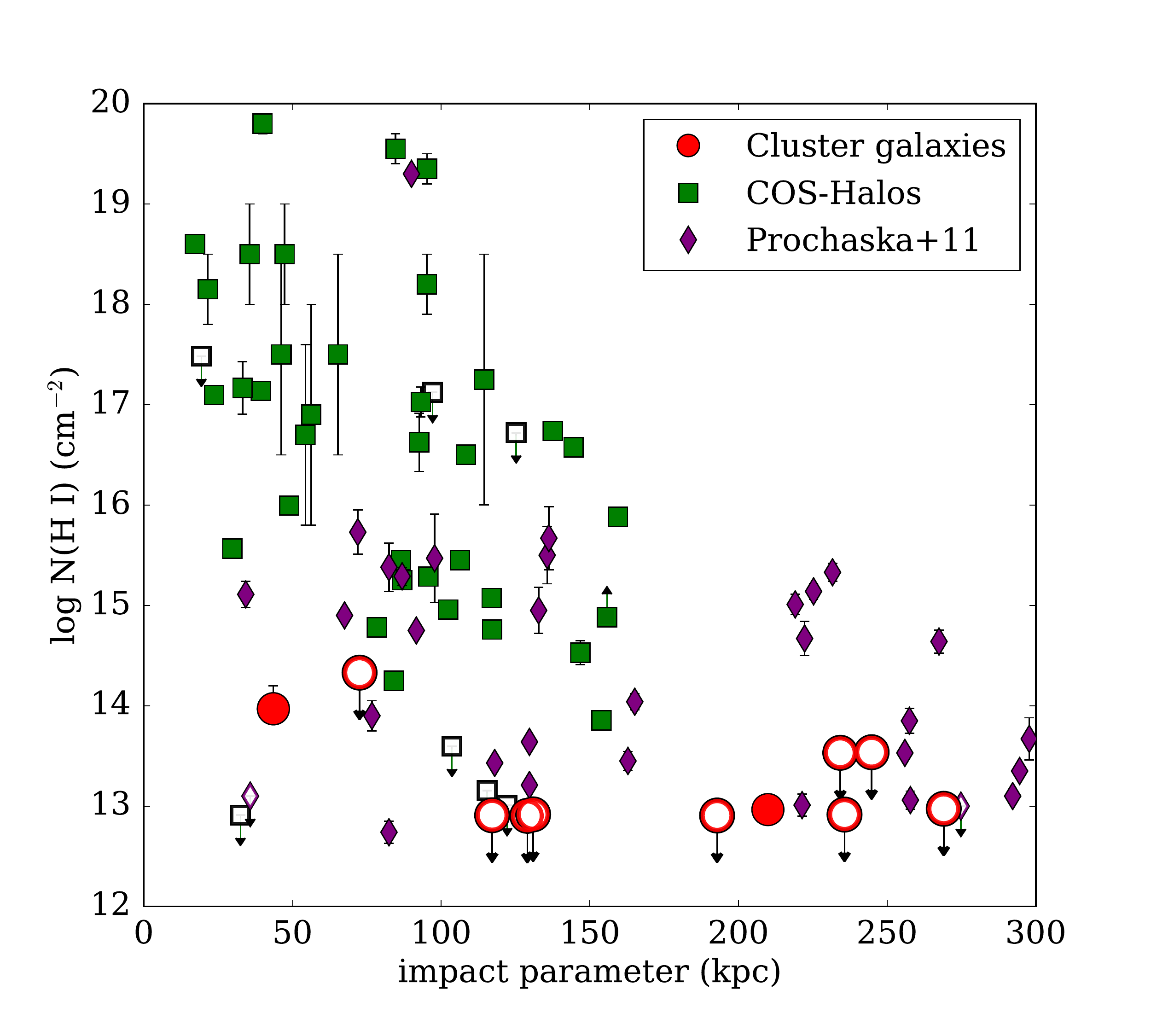}
\caption{The \hone\ column densities measured in our cluster galaxies' CGM compared with field galaxy CGM measurements from the literature.  The red circles denote cluster galaxies from our sample, while the green and purple points correspond to CGM measurements of field galaxies. All upper limits (open symbols) are $3\sigma$. The filled symbol at $\rho=$ 44 kpc corresponds to the unambiguous plus possibly blended \hone\ absorption for the galaxy in A2246.  The column densities and galaxy information for our cluster galaxy sample may be found in Table \ref{tab:CGMmeas}.}
\label{fig:HIprofile}
\end{figure}

\begin{figure} 
\includegraphics[width=1\linewidth]{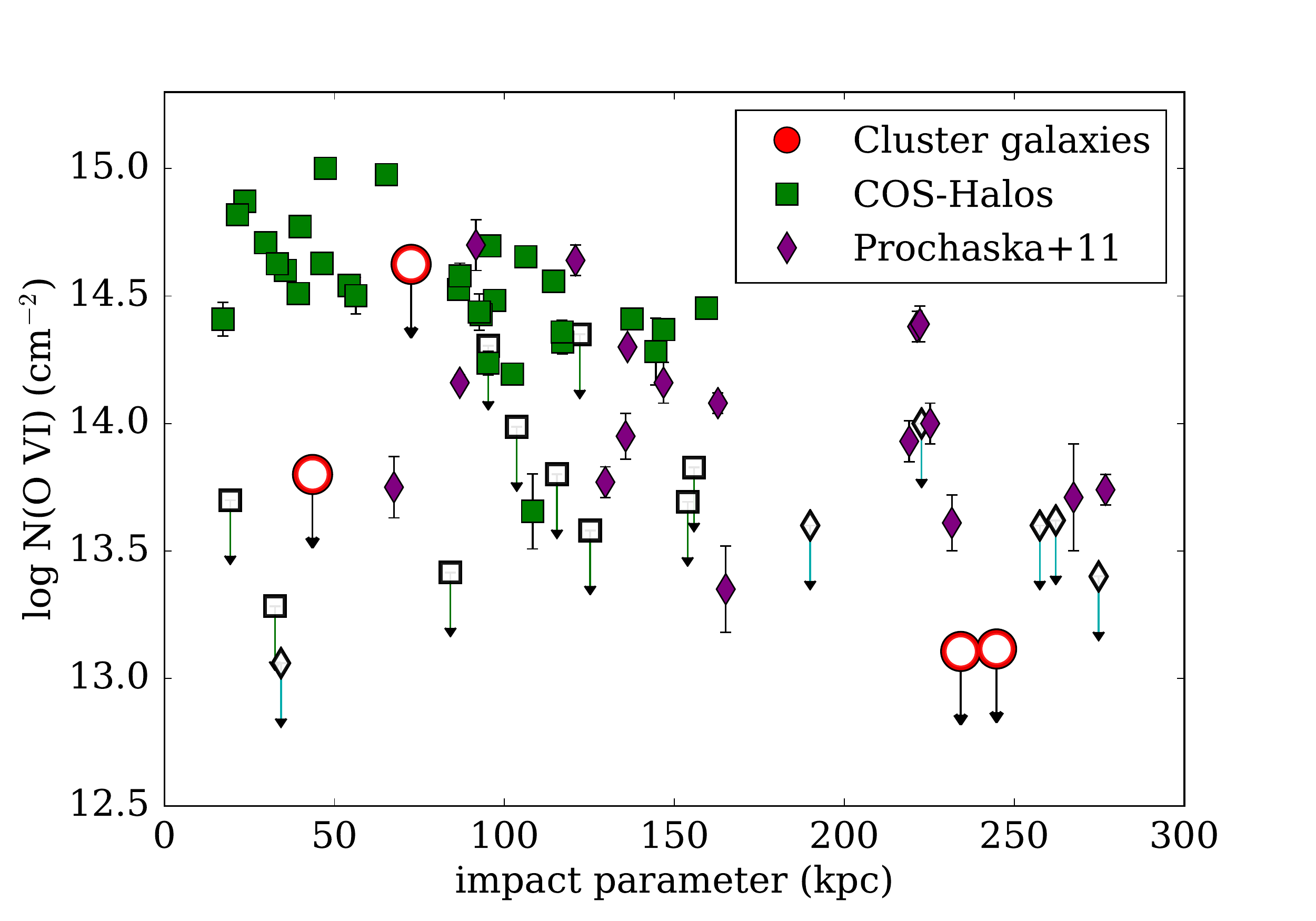}
\caption{The \osix\ column densities measured in our cluster galaxies' CGM compared with field galaxy CGM measurements from the literature.  The red circles denote cluster galaxies from our sample, while the green and purple points correspond to CGM measurements of field galaxies. All upper limits (open symbols) are $3\sigma$. The column densities and galaxy information for our cluster galaxy sample may be found in Table \ref{tab:CGMmeas}.}
\label{fig:OVIprofile}
\end{figure}

We also searched for \osix\ absorption in the CGM of these same galaxies for which the $\lambda \lambda$ 1032, 1038 \AA\ \osix\ transitions fall within the COS bandpass at the galaxy redshifts.  While \hone\ is covered for all five cluster systems considered in this work, \osix\ is only redshifted into the bandpass for four of them, producing a CGM \osix\ sample of only four galaxies.  A2246 contains a galaxy at $z=0.22849$ and impact parameter of $\rho = 43$ kpc; \gmb\ contains one at $z = 0.45599$ and $\rho=72$ kpc.  The QSO spectrum corresponding to A2246 and \gmb\ (HS1700+6416; Figures \ref{fig:Stackplot3} and \ref{fig:Stackplot4}), being of a high-redshift QSO sightline that is heavily populated with absorption features, shows several interloping lines from different redshifts but no unambiguous \osix\ features within $\pm 400$ km s$^{-1}$ of the galaxy redshifts.  The combined constraints from both lines of the doublet only allow very weak \osix\ lines near the redshift of the A2246 galaxy, and none of the data provide compelling evidence of any \osix. We establish an upper limit on N(\osix) for the A2246 galaxy by fitting Voigt profiles to the complexes of absorption at the would-be locations of \osix\ assuming that the \osix\ and \hone\ components potentially associated with this galaxy have the same velocity structure.  For the \gmb\ galaxy, we find that a substantial amount of \osix\ could be hidden in nearby blended features, but there is no \hone\ at the velocities of those features.  To derive a conservative upper limit, we allowed the possibility that the \osix\ and \hone\ are decoupled in this environment, and we used component fitting to determine the maximal amount of \osix\ that could be hidden in those blends. The resulting Voigt profile model that provides this upper limit is indicated by the red line in the lower two panels of Figure \ref{fig:Stackplot4}, and we report the sum of column densities from these models as the upper limit.  A1095 is the remaining cluster with galaxies $\rho < 300$ kpc from the sightline and also contains no unambiguous components of \osix.  These data are plotted alongside the COS-Halos \citep{Tumlinson:2011kx} and \citet{Prochaska:2011yq} samples in Figure \ref{fig:OVIprofile}.

\subsection{The Warm-hot Gas Content of Galaxy Clusters}
\label{sec:WaHoGa}
As found by G16, the baryonic mass within $r_{500}$ accounted for by stars and X-ray emitting gas is approximately 9\% and 10\% of $M_{500}$ in the subclusters of A1926 and A1095, respectively.  These measurements fall short of the Universal baryon fraction, 16.6\% \citep{Bennett:2013kx}, potentially signaling that the clusters may contain some fraction of their gas in some other phase.  Most of our QSO sightlines piercing the ICM of these clusters fall beyond $r_{500}$, but the impact parameter of A1095W is approximately equal to $r_{500}$.  The spatial coincidence of our detected X-ray emission and the QSO spectra enable a unique opportunity to directly constrain the relative contributions of gas in the hot phase from X-rays and in the warm-hot phase from the QSO data, particularly in the poorly constrained regions beyond $r_{500}$.

\subsubsection{Diagnostic: \osix}

The $\lambda \lambda$ 1032,1038 \AA\ \osix\ doublet is the most frequently used tracer of $10^{5-6}$ K gas, and we designed our experiment such that the COS G130M grating would cover the \osix\ doublet at the redshifts of A1095, A1926, and A2246.  \osix\ shifts into the COS/G130M bandpass at approximately $z\sim0.1$ and is therefore covered by our data at the redshift of \gmb\ but not for \maxb.  As seen in Figures \ref{fig:Stackplot1} - \ref{fig:Stackplot5}, no significant, unambiguous \osix\ features are detected within 1500 km s$^{-1}$ of the cluster redshifts.

\subsubsection{Diagnostic: Broad \hone}
\label{sec:broadHI}
In addition to \osix, warm-hot gas may also be detected as thermally broadened spectral features, most notably \hone\ \lya\ \citep[broad \lya\ or BLA; e.g.,][]{Lehner:2007lr}.  The line broadening of the Voigt profile is parameterized by the Doppler $b$ value.  In general, the $b$ value contains both thermal and nonthermal contributions.   The thermal contribution is simply related to the gas temperature through the kinetic energy of the gas particles:

\beq
b_{th} = \sqrt{\frac{2kT}{m}} = 0.129\sqrt{\frac{T}{A}} ~~\rm{km~s^{-1}}
\eeq

\noindent where k is the Boltzmann constant, T is the gas temperature, and m and A are the particle mass and atomic mass number, respectively, of the species responsible for the absorption.  The rightmost expression corresponds to temperatures measured in Kelvin and yields units of km s$^{-1}$.  Therefore, gas at the temperatures of interest ($10^5 - 10^6$ K) has corresponding $b$ values of 40-150 km s$^{-1}$ for hydrogen.  Non-thermal contributions may arise from turbulence or bulk motions; furthermore, unresolved blends of narrow components can mimic broad features. For our conversions between Doppler $b$ and temperature, we assume a nonthermal contribution of $30\%$ to the line width from turbulence as measured by \citet{Hitomi:2016lr} in the Perseus cluster\footnote{The $Hitomi$ satellite obtained a high-resolution spectrum in the Perseus cluster ICM, enabling simultaneous measurements of the plasma temperature and widths of emission lines from Fe.   From these Fe lines, they measure a line of sight velocity dispersion due to turbulence of $164 \pm 10$ km s$^{-1}$, having removed a thermal contribution of 80 km s$^{-1}$.  Scaling this thermal velocity to that expected for hydrogen nuclei (the dominant species) via the atomic masses of Fe and H (cf. Equation 1) yields a thermal velocity of $\sim600$ km s$^{-1}$, of which the turbulent contribution $\sim30 \%$.}. Despite the intrinsic strength of \lya, line broadening will make gas of a given column density more difficult to detect, hampered by both the decreased optical depth per resolution element and uncertainty of the continuum placement. 

Our COS data cover \lya\ for A1095 and A1926 and \lyb\ for A2246.  For A2246, \lya\ is covered by the archival STIS spectrum of HS1700+6416, albeit at much lower S/N than the COS data for the other sightlines.  We detect BLA features in the QSO spectrum probing one cluster: A1926 (Figure \ref{fig:Stackplot2}).  The features are well-fit by Voigt profiles at velocities $v=-937$, -736, and 1061 km s$^{-1}$ relative to the adopted cluster redshift.   The resulting $b$ values for these components are $69 \pm 9$, $61 \pm 24$, and $71\pm23$ km s$^{-1}$, which correspond to temperatures of log T = 5.1, 5.0, and 5.2 K, respectively.

Because the line broadening can be significant at the temperatures we are interested in probing with the QSO spectroscopy, we wish to constrain, as a function of temperature, the absorbers to which our study is sensitive.  In practice, determining the statistical significance of a spectral feature involves fitting a continuum, measuring the equivalent width of the feature and the uncertainty in this measurement, and taking the ratio of the equivalent width and its uncertainty.  Statistical uncertainties arise from both the optical depth of the line relative to the noise of the data and the shape of the fitted continuum.  Fitting a continuum requires choosing line-free segments of the spectrum blueward and redward of the desired feature to be measured, introducing considerable systematic uncertainty given the combination of low-frequency undulations that are empirically known to exist in QSO spectra and the possible presence of substantially broadened absorption lines.  While the uncertainty in the continuum can be characterized given sufficient knowledge of line-free regions \citep{Sembach:1992rt}, the systematic uncertainty from the probably false assumption that these regions are free of lines is extremely difficult to quantify. 

However, the parameter space of detectable absorbers can be somewhat constrained using the idealized scenario where a) the continuum is well behaved (i.e., free of undulations in the region of the spectrum where the BLA is to be measured) and b) the line-free regions are clearly selected on either side of the BLA.  Indeed, calculating the detection significance of BLAs as a function of column density and $b$ value under these idealized circumstances will identify a class of absorbers that are completely undetectable at a given S/N level.  To do so, we conduct a simulation experiment as follows: First, we generate Voigt profiles over a grid of column densities, log N(\hone) = $11 - 16~	\cmt$, and $b$ values that correspond to temperatures log T $\sim 4-6 $ K.  Then, we inject each absorber onto a flat continuum, convolve with the COS line spread function, and add Gaussian noise commensurate with the S/N levels in our data at the locations of \lya\ lines with redshifts equal to the galaxy clusters they probe. Using the formalism of \citet{Sembach:1992rt}, we fit a continuum declaring two line-free regions on either side of the injected absorption line (with centroid \lamnot): between the blue edge of the simulated `spectrum' (\lamnot\ $- 10$ \AA) and \lamnot $-\frac{2b}{c}$ and between \lamnot $+\frac{2b}{c}$ and the red edge of the spectrum, where $c$ is the speed of light.  Finally, we measure the equivalent width of the line ($W$) and the uncertainties from both the noise in the flux ($\sigma_{W_f}$) and that due to the uncertainty from the continuum placement \citep[$\sigma_{W_c}$;][]{Sembach:1992rt}.  For each simulated absorber, the detection significance is then

\beq
S = \frac{W}{\sqrt{\sigma_{W_f}^2 + \sigma_{W_c}^2}}
\eeq  

\noindent where $W$, $\sigma_{W_f}$, and $\sigma_{W_c}$ are all measured over velocity widths of $dv \in [-b/2,+b/2]$.

Figure \ref{fig:Detectoplot} shows the detection significance of simulated absorbers across the full grid of N(\hone) and $b$ values.  In absorption line studies, typical thresholds for detection are $3-5 \sigma$, and Figure \ref{fig:Detectoplot} shows that the absorber column densities that may be detected \emph{even given the idealized continuum placement conditions} increases by $\sim$ 0.5 dex across the range of $b$ values (corresponding to log T $\sim 4-6 $ K assuming purely thermal line broadening).  Thus, in terms of column density, the  \emph{maximum} sensitivity drops by more than half across this temperature range.  Furthermore, the uncertainty in continuum placement will only exacerbate the decreased sensitivity to higher temperature gas.  Lastly, we assessed the likelihood of noise features mimicking BLA absorption through a similar experiment but without injecting synthetic line profiles.  We confirm that the false positive rate of purely Gaussian noise producing significant BLA-like features is $<0.01\%$; therefore, the dominant source of error is most likely the continuum placement uncertainty.

\begin{figure} 
\includegraphics[width=1.1\linewidth]{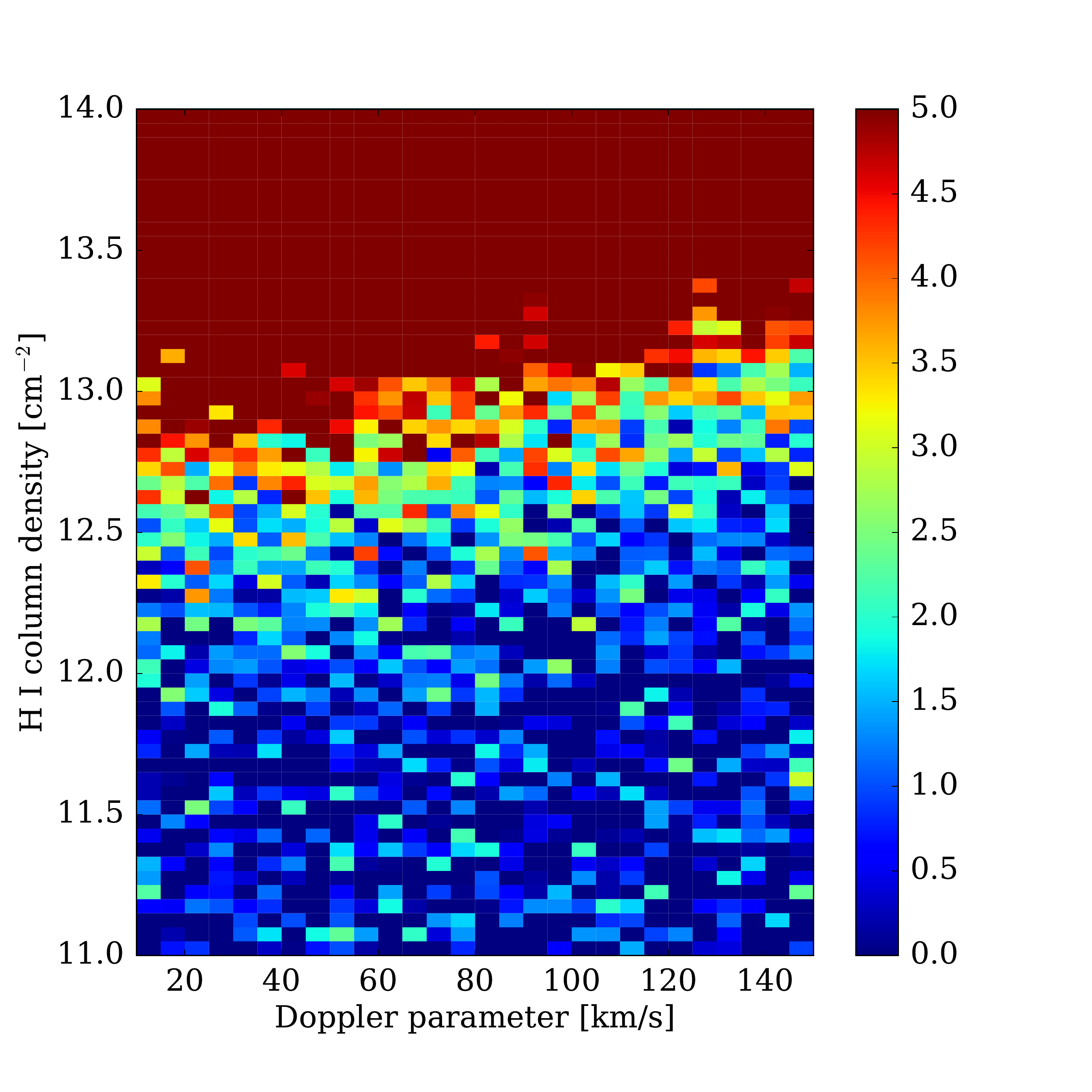}
\caption{Detection significance of idealized simulated \lya\ absorbers (see colour bar on right) as a function of N(\hone) and $b$ value where the noise of the simulated spectrum is commensurate with those probing A1095 and A1926 in our sample.  For a given column density N(\hone) $> 10^{12.5}~\cmt$, absorption features become less detectable with increasing Doppler parameters.  Adopting a significance threshold of $3-5 \sigma$, the column density sensitivity decreases by 0.5 dex over the range of Doppler parameters shown, which correspond to temperatures log T $\sim 4-6$ K assuming pure thermal broadening.}
\label{fig:Detectoplot}
\end{figure}

\subsubsection{The relative baryon contributions of the hot and warm-hot phases}
We now quantify the amount of gas in the $10^{5-6}$ K warm-hot phase along the line of sight probing A1926 to place in context with that derived by G16 for the $>10^7$ K hot phase.  Using the column densities of \hone\ measured by Voigt profile fitting, we convert to total H column density (N(H$^+$) + N(\hone)) using an ionization correction. We employ here the \citet{Oppenheimer:2013mz} ionization models evaluated at a metallicity of Z = 0.3 Z$_\odot$ and total hydrogen density of $n_H = 10^{-4} \rm{cm}^{-3}$.  These models feature photoionization equilibrium (although we also experimented with their nonequilibrium photoionization models and find that this added effect did not change our results) and assume that the gas is cooling from an initial temperature of $10^7$ K. We adopt the models run with a \citet{Haardt:2012fj} ionizing photon background evaluated at $z=0.2$.  The ionization corrections we derived were insensitive to these model parameter choices, deviating by $\sim 0.05$ dex.

\begin{figure} 
\includegraphics[width=1.\linewidth]{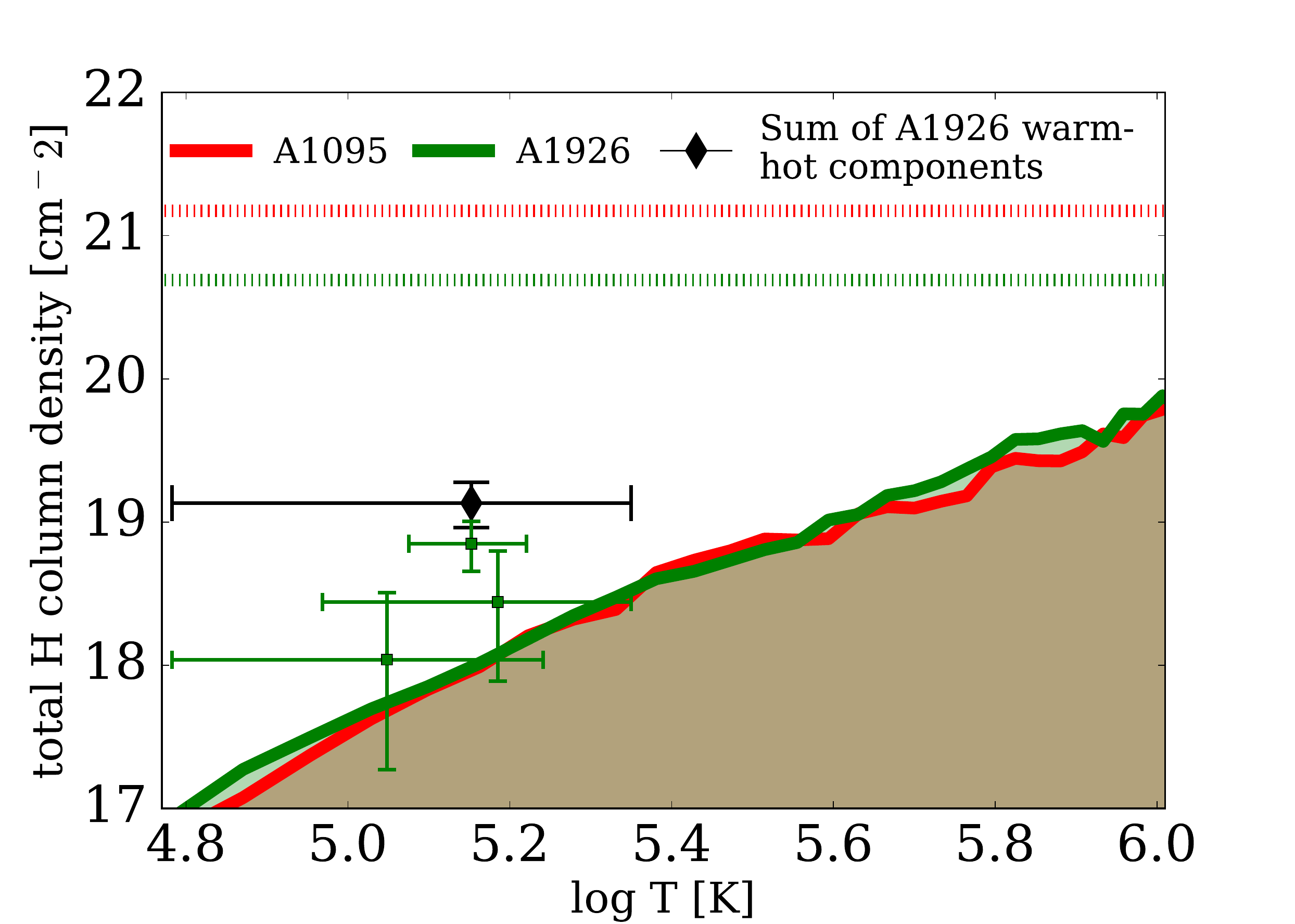}
\caption{The contributions of the warm-hot and hot gas phases to the total hydrogen column density (N(H) = N(\hone + H~\textsc{ii})) along the QSO sightlines probing A1095 and A1926 as a function of temperature.  The red and green dashed lines mark the $>10^7$ K hot gas contributions to N(H) for A1095 and A1926, respectively, measured at the locations of the QSO sightlines by G16.  The green points with error bars correspond to the 3 \lya\ components detected near the redshift of A1926 in the UV spectrum of 2MASS J1431258+244220.  The red and green lines and shaded regions below them depict the limits on N(H) from potentially detectable (but undetected in the actual data) BLA absorbers given the idealized conditions described in Section \ref{sec:broadHI}.  Given the idealized nature of our experiment, the true upper limits would be higher (see Section \ref{sec:broadHI}), thus potentially hiding significant columns of warm-hot gas.  The black diamond with error bars shows the summed contribution from the BLAs detected near A1926 with derived temperatures $T =10^{5-6}$~K.  }
\label{fig:WaHoGaMass}
\end{figure}

The \citet{Oppenheimer:2013mz} models provide ion fractions as function of temperature for the ionization states of several species.  We calculate N(H) as 

\beq
N(H) = \frac{\rm N(\hone)}{\chi_{\hone}(T,n_H,Z,z)}
\eeq

\noindent where $\chi_{\hone}$ is the ionization ratio n$_{\hone}$/n$_H$ and the other parameters are set as above.   

In Figure \ref{fig:WaHoGaMass}, we present the total column density of hydrogen resulting from this calculation for the three broad \lya\ components associated with A1926 and reported in Section \ref{sec:broadHI} (green points with error bars), the contribution measured from the X-ray emitting gas in G16 (dashed lines), and the limits placed on the total H column using the idealized simulated absorption (green and red solid lines and shaded regions) as described in the previous section.  The bold, black diamond and error bars represent the sum of the three components whose Doppler $b$-values suggest gas temperatures of 10$^{5-6}$ K, N(H) $\sim10^{19}~\cmt$.

By comparison, the total warm-hot gas detected in the BLA absorbers associated with A1926 (black marker in Figure \ref{fig:WaHoGaMass}) amounts to approximately 3\% of that in the hotter phase traced by X-rays.      We note that the warm-hot material may also be photoionized by the radiation emitted by the nearby $10^7$ K gas, which would further decrease $\chi_{\hone}$ and increase the derived N(H) in the warm-hot phase, although the photoelectric cross-sections of $\sim 1$ keV photons are very small. Nevertheless, these photon contributions are not included in our modeling.  However, the  Because of the continuum and flux uncertainties described in Section \ref{sec:broadHI}, more warm-hot material may reside in undetectable lower-column density clouds.  In fact, we caution that the limits represented by shaded regions in the Figure \ref{fig:WaHoGaMass} underestimate the true limits due to the idealized nature of our absorption experiment.  Particularly with the systematic continuum placement uncertainty involved in choosing line-free regions to fit continuua, the detectability of weaker BLAs in actual QSO spectra can also depend on the spectral region of interest being fortuitously flat.  

These uncertainties notwithstanding, the warm-hot phase may only represent a significant contribution to the baryon budget in the outer regions of A1926 if a large mass of the material is actually at log $T>5.5$ K, with broad lines that are extremely difficult to detect.  We emphasize that the BLA features we detect lie at velocities $\Delta v \sim \pm1000$ km s$^{-1}$ relative to the cluster redshift.  While these velocities are consistent with the velocity dispersion of an $M_{200}>10^{14}$ \msun\ cluster, they would correspond to distances of $>14$ Mpc if due to pure Hubble flow.  Assuming that these absorbers are associated with A1926, the velocities may be an important clue that the absorbing gas is falling into the cluster.  We return to this point in Section \ref{sec:discussBLA} for further discussion.  

\subsection{Absorption as a function of clustocentric impact parameter}
We now examine our absorption data with regard to the centre of each galaxy cluster probed by our QSO sightline data.  X-ray measurements clearly reveal that the hot gas temperature, density, and metallicity vary widely from the central cluster regions outward.  Indeed, \citet{Yoon:2012yu} and \citet{Yoon:2017aa} conclude that the $\sim10^4$ K gas probed by their QSO sightlines piercing the Virgo and Coma cluster regions is deficient within their virial radii and at small velocity separations to the cluster redshifts.  Thus, the hot X-ray emitting gas may dominate the inner ICM, while infalling gas directly from the IGM or within infalling bound groups or structures may comprise the absorbers they detect in \lya.  More recently, \citet{Muzahid:2017lr} observed sightlines on the outskirts of three Sunyaev-Zel'dovich effect-selected clusters and detected strong \hone\ absorbers (log N(\hone) $>$ 16.5 $\cmt$) at similar redshifts to their galaxy clusters in three sightlines.

Taking a clustocentric perspective, Figures \ref{fig:ClustoKpc} and \ref{fig:ClustoVir} show the equivalent width of \lya\ or limits thereof as a function of projected impact parameter to the centre of each cluster probed by our QSO sightlines.  For each cluster system, the impact parameter adopted is relative to the X-ray centroids found by G16 and WW14; for systems where two X-ray centroids were measured, we adopted the one with the smallest impact parameter.  As described in Section \ref{sec:resultsCGM} above, we measure both the unambiguous absorbers and possibly blended absorption coincident with the cluster redshifts (adopting limits of $| dv | < 2000$ km s$^{-1}$).  For the \gmb\ system, at redshift $z = 0.452$, we estimate the \lya\ equivalent width based on the measured column density of the \lyg\ line, the lowest Lyman series line covered by our COS data.  Uncertainties shown in these figures are purely statistical and do not include those due to the continuum placement. Figure \ref{fig:ClustoVir} presents the absorber strength as a function of impact parameter but normalized to \rtwo.

\begin{figure} 
\includegraphics[width=1.1\linewidth]{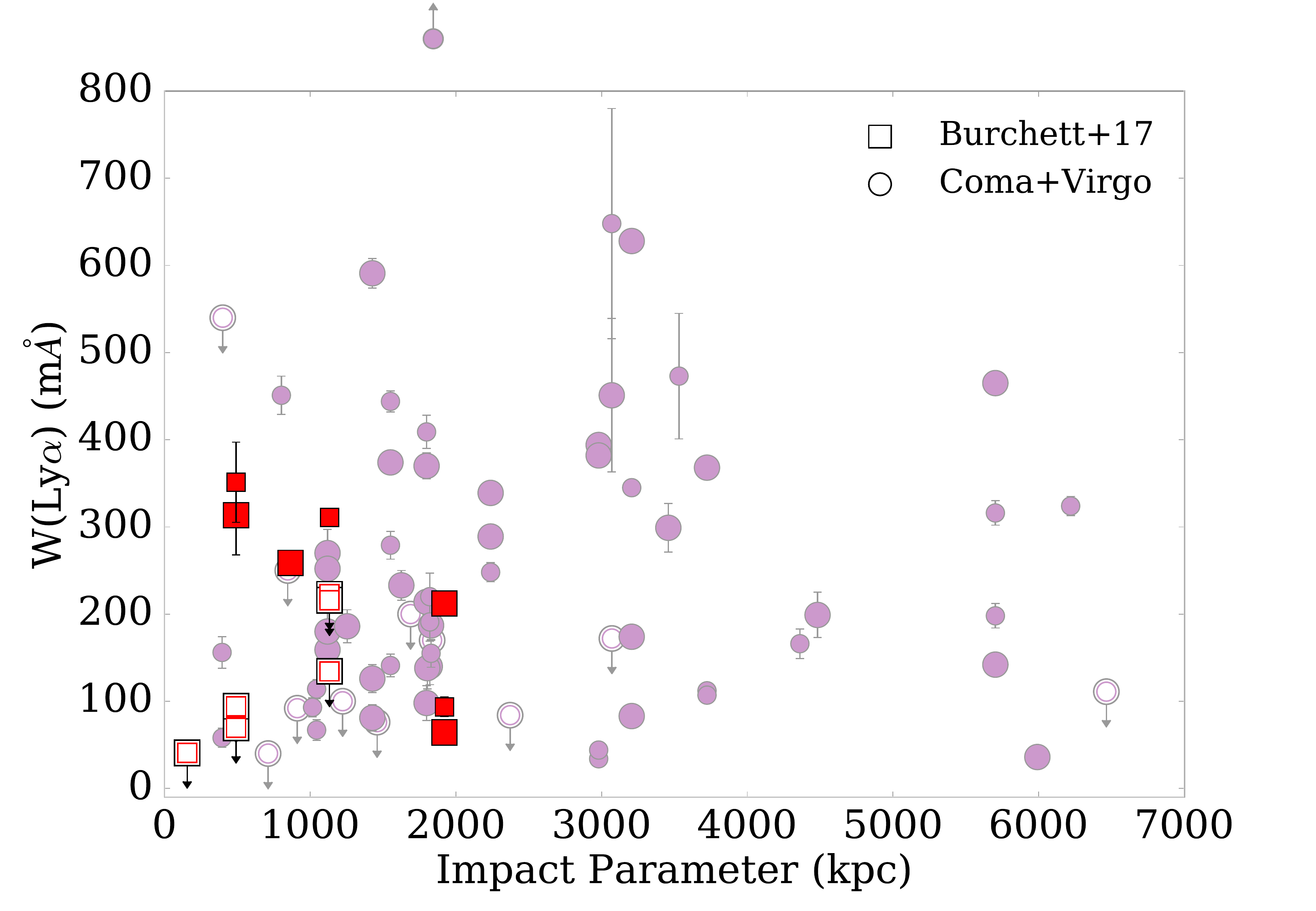}
\caption{\hone\ absorption as a function of projected distance between QSO sightlines and galaxy clusters with absorber-cluster velocity separations $<2000$ km s$^{-1}$.  The faint purple circles represent data from \citet{Yoon:2012yu} and \citet{Yoon:2017aa} probing the Virgo and Coma clusters, and the red squares denote the absorbers and limits from our survey.  Filled symbols represent unambiguous detections of \hone, while open symbols denote upper limits on \hone\ absorption (for our data, these include where potential components may be blended with lines from other redshifts).  Symbols are sized according to their absorber-cluster velocity separations, where larger symbols indicate $|dv| < 1000$ km s$^{-1}$.  The purple square plotted above the frame of the plot corresponds to a sub-damped \lya\ system first discovered by \citet{Tripp:2005lr} with W(\lya)$\sim2240$ m\AA.  Note that the three upper limits from our data with impact parameter $\sim$ 1100 kpc (probing \gmb) are estimated from limits on \lyg\ absorption because the cluster redshift places \lya\ and \lyb\ beyond the coverage of our COS spectrum.  Consistent with a scenario of increased \hone\ absorption beyond the inner cluster regions, our sightline with the largest clustocentric impact parameter ($\sim2$ Mpc) shows 3 components of \lya, while the 4 sightlines at smaller projected distances show only 4 unambiguous \hone\ components among them, with 2 having $|dv| > 1500$ km s$^{-1}$.}
\label{fig:ClustoKpc}
\end{figure}

\begin{figure} 
\includegraphics[width=1.1\linewidth]{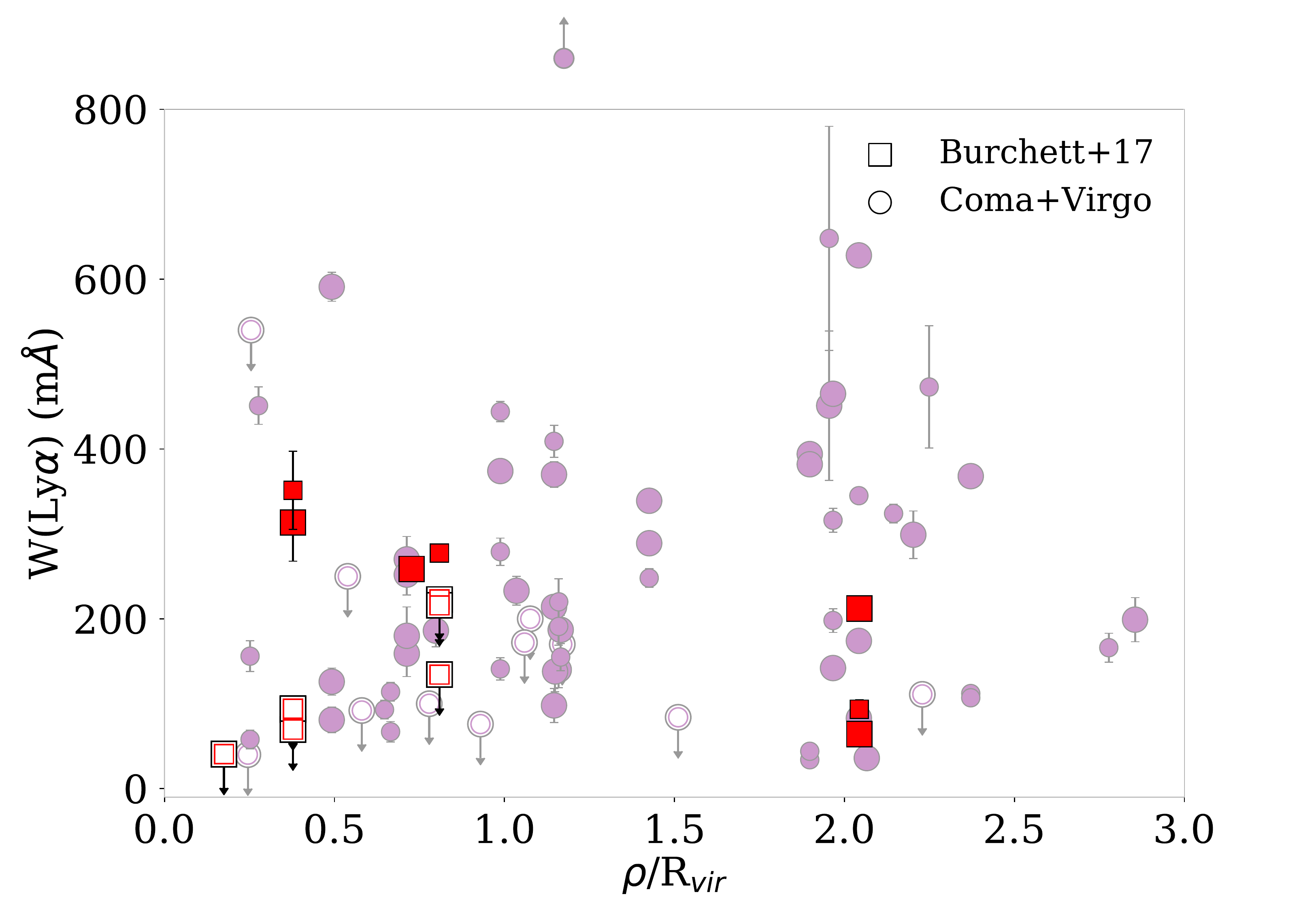}
\caption{Similar to Figure \ref{fig:ClustoKpc} but with impact parameters normalized to the virial radii of each cluster.}
\label{fig:ClustoVir}
\end{figure}

As seen in both figures, the cluster showing the greatest number of \hone\ absorption components is the cluster probed at the furthest impact parameter (A1926 at $\sim 1900$ kpc).  Of the four clusters probed at smaller impact parameters, three show unambiguous \hone\ absorbers but only two have absorbers at $|dv| < 1000$ km s$^{-1}$ from the cluster redshift (plotted with larger squares).  We note that one of these clusters (A1095) is highly dynamically unrelaxed (see Figure \ref{fig:XrayMapMontage}), which may enable unvirialized, cooler gas to reside in this environment.  In Figure \ref{fig:ClustoVir}, we have normalized by the cluster virial radii: A1926, with three absorption components, is probed at $>2$ \rtwo; the remaining four sightlines with four absorption components among them, have impact parameters within 1 \rtwo.  As shown by the symbol sizes, the two strongest components at $\rho < $ \rtwo\ have relatively large velocity separations from the cluster redshift.  In fact, the absorber-cluster separations for these two components have $|dv| > 1500$ km s$^{-1}$, greater than the cluster velocity dispersions and therefore possibly unrelated to the clusters themselves.  Plotted in faint purple in Figures \ref{fig:ClustoKpc} and \ref{fig:ClustoVir} are the combined data from \citet{Yoon:2012yu} and \citet{Yoon:2017aa} targeting the Virgo and Coma clusters, also within the absorber-cluster relative velocity range $| dv | < 2000$ km s$^{-1}$. These Virgo and Coma sightlines show a higher covering fraction of \hone\ in the outer cluster regions than within 1 \rvir. Interestingly, as reported by \citet{Yoon:2017aa}, this difference is driven by Virgo; Coma shows a similar covering fraction within and beyond 1 \rvir.  A dearth of \hone\ in central cluster regions relative to their outskirts may be further corroborated according to \citet{Muzahid:2017lr}, as they detected 3 strong \hone\ absorbers at velocities consistent with the uncertainties of the photometric redshifts beyond the virial radii of their clusters.  \citet{Muzahid:2017lr} suggest that their data provide evidence for an abundant \hone\ reservoir in the outskirts of galaxy clusters.

An excess of cool $10^4$ K gas in the outer regions of clusters relative to the hot inner regions characterized by dense, hot $>10^7$ K gas would hardly be surprising, but (1) confirmation of this excess with absorbers more unambiguously associated with the targeted clusters and (2) the nature of this gas (infalling material from the IGM, the CGM of cluster galaxies, etc.) beg further investigation. The large uncertainties in the cluster photometric redshifts and large cluster-absorber velocity separations ($>2000$ km s$^{-1}$) of the \citet{Muzahid:2017lr} complicate points (1) and (2), although indirect arguments from absorber statistics (d$\mathcal{N}$/d$z$) help bolster their claim of association between their absorbers and clusters.  \citet{Yoon:2017aa} show that the covering fraction increases for both Virgo and Coma when including wider velocity ranges about the cluster redshift ($2 \sigma_{cl}$}.  The three absorption components we detect in the sightline probing A1926 all have spectroscopically constrained velocity separations of $|dv| \lesssim 1000$, but the components are still separated by $\gtrsim 700$ km s$^{-1}$, which could correspond to distances of $>10$ Mpc at Hubble flow velocities; separations of $2000$ km s$^{-1}$ may correspond to Hubble flow distances of $>55$ Mpc.  Larger samples of absorbers on the outskirts of spectroscopically measured clusters with small velocity separations will help confirm an \hone\ excess, and surveys of the cluster member galaxies will enable further CGM investigation  but also help rule out galaxy associations and reveal an intracluster or infalling intragroup nature to the traced gas.  Achieving these goals will require ambitious observational investments of current space-based and ground-based facilities.  

\section{Discussion}
\label{sec:discussion}

\begin{figure} 
\includegraphics[width=1.15\linewidth]{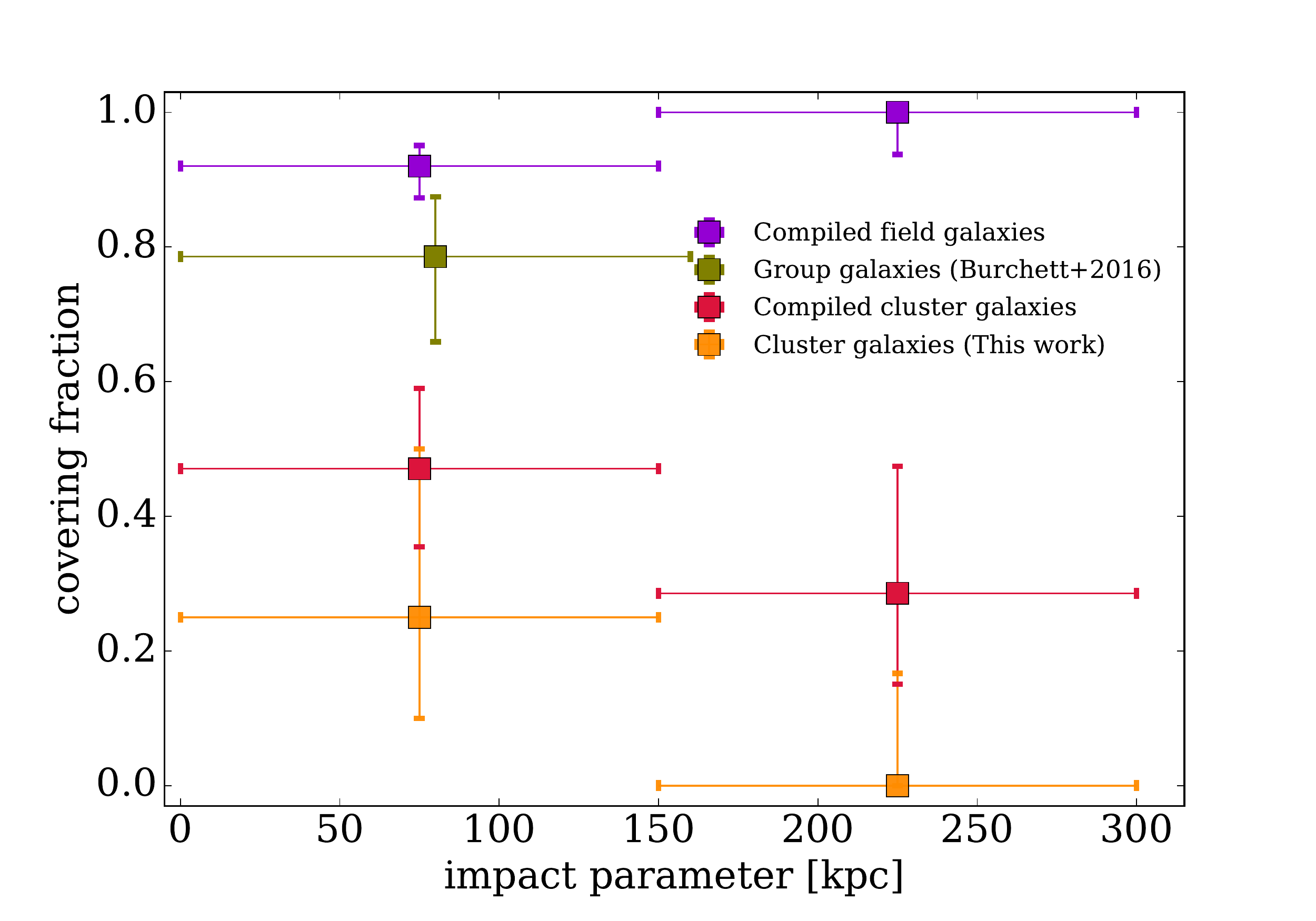}
\caption{Covering fractions of \hone\ in the CGM, calculated in bins of impact parameter indicated by the horizontal lines, of galaxies across various environments.  Shown are a field galaxy sample (purple) combining those of \citet{Tumlinson:2013cr}, and \citet{Prochaska:2011yq}, galaxies in group environments (green) from \citet{Burchett:2015aa}, a cluster sample (red) combining this work and that of \citet{Yoon:2013kq}, and the data from this work alone (orange).  For comparison, we have imposed the same 100 m\AA\ detection threshold as \citet{Yoon:2013kq} for our covering fraction measurement.  The cluster galaxies show markedly lower covering fractions than those of field galaxies.  However, galaxies in groups may lie between these two populations, perhaps having undergone intermediary levels of the stripping/heating processes that remove the \hone\ from galaxy halos.}
\label{fig:CovFrac}
\end{figure}

\subsection{A continuum of environmental impacts on the CGM}  
In Section \ref{sec:resultsCGM}, we have presented evidence that the CGM of galaxies in our cluster sample has been stripped or has become sufficiently heated and ionized such that the column densities of neutral hydrogen fall well below measurements reported for galaxies in less dense environments.  Previous studies have also reported environmental effects on the CGM, and we now place our results in context with their findings.

\citet{Yoon:2012yu} and \citet{Yoon:2013kq} employed a sample of QSO sightlines that pierce the nearby Virgo cluster and are perhaps the most directly comparable studies to ours.  \citet{Yoon:2012yu} reported statistics of \lya\ absorbers relative to the cluster as a whole and find that the covering fractions of strong \lya\ absorbers ($>100$ m\AA) are lower within 1 \rvir\ of the cluster ($60^{+16}_{-13}\%$) than from 1-2 \rvir\ ($100_{-14}\%$), suggesting that the physical conditions inside the cluster virial radius are such that the neutral gas content is suppressed, likely shock-heated and ionized. \citet{Yoon:2013kq} studied associations between \lya\ absorbers and individual cluster galaxies and report significantly lower covering fractions of \lya\ around individual galaxies within 1 \rvir\ of Virgo than galaxies in the field.  Using a similar velocity cut as we used in Section \ref{sec:resultsCGM} and that of \citet{Prochaska:2011yq}, they find $f_c = 39^{+14}_{-12}\%$ for their cluster galaxies within 1 \rvir\ of Virgo.  Their result is within the errors of the covering fraction we report in Section \ref{sec:resultsCGM} ($f$(\hone) = $25^{+25}_{-15}\%$).  In Figure \ref{fig:CovFrac}, we show the covering fractions both from our sample (orange) and as a result of combining the \citet{Yoon:2013kq} sample with ours (red) in two bins of impact parameter.  Here, we have imposed the same 100 m\AA\ equivalent width limit as \citet{Yoon:2013kq}; their galaxy-absorber velocity association criterion ($|\Delta v| < 400$ km s$^{-1}$) matches that employed here and by  \citet{Prochaska:2011yq}. While our measurements are not discrepant with \citet{Yoon:2013kq} at any statistical significance, one may expect their sample to have an intrinsically higher covering fraction due partly to sample selection: they adopt $R_{100}$ as \rvir\ and include galaxies within this projected boundary.  All of our associated galaxy/sightline associations lie projected within \rtwo.  

However, another intriguing comparison arises with the CGM of galaxies in groups.  Figure \ref{fig:CovFrac} shows the covering fraction of $\hone$ associated with a sample of group galaxies from \citet[][plotted in green]{Burchett:2015aa}.  The group galaxies exhibit an \hone\ covering fraction intermediate to the cluster and field samples, suggesting that the processes removing the CGM \hone\ from cluster galaxies set in with progressively higher efficiency as the environmental density increases. \citet{Burchett:2015aa} used a fixed-aperture density metric to quantify galaxy environment (see their Figure 6), and we are including the galaxies from their sample with 5 or more $L>0.1~L*$ neighbours within 1.5 Mpc and 1000 km s$^{-1}$ as `group galaxies'.  According to the assessment by \citet{Muldrew:2012qy} of several environmental metrics, this parameter choice should conservatively select group-sized halos ($\mhaloeq > 10^{13}~\msuneq$) in the sense that many lower-mass halos are also likely to be included.  Given that more isolated galaxies have near-unity \hone\ covering fractions, the inclusion of galaxies in less-massive halos should increase the covering fraction; indeed the non-detections of \hone\ in the \citet{Burchett:2015aa} sample all occur at densities $>10$ galaxies per 1.5 Mpc aperture, and cutting at higher densities than we have chosen would yield an even lower \hone\ covering fraction and thus all the more discrepant with that of the field galaxies.   At impact parameters $<350$ kpc, \citet{Wakker:2009fr} also found a lower covering fraction of \lya\ absorbers around group galaxies than field galaxies, $61\pm17\%$ and $100_{-37}\%$, respectively.  In addition to the differences in absorber-galaxy impact parameters, their group identification/membership scheme differs from that employed by \citet{Burchett:2015aa} and we caution against direct quantitative comparison of the two results.  However, the \citet{Wakker:2009fr} result qualitatively agrees that the \lya\ covering fraction lies between that of clusters and field galaxies. 

Few instances of ram pressure acting in groups have been observed directly \citep[e.g.,][]{Rasmussen:2006uq,Freeland:2011aa}.  Several studies, observational and theoretical, have examined galaxy quenching in groups to better constrain the quenching mechanisms \citep[e.g.,][]{Fillingham:2016aa,Rasmussen:2012aa,Kawata:2008aa}.  The efficiency and relative importance of ram-pressure stripping to directly remove galaxies' cold disk gas within galaxy groups remains unclear. The question of what disturbs and even quenches group galaxies is further complicated by the variety in dynamical states of the groups and merger/harassment history of the galaxies therein.  However, as suggested by Figure \ref{fig:CovFrac}, the CGM is increasingly stripped and/or heated as the environmental density increases, which may quench the host galaxies on longer timescales (i.e., strangulation) even if the cold gas in the disk is not directly stripped.  The unparalleled sensitivity and kinematic information provided by QSO absorption line spectroscopy can and should be leveraged to further diagnose these physical processes across the full environmental spectrum.  However, substantially larger samples and uniformly identifying and quantifying mass, richness, etc., for groups and clusters will be critical.

\subsection{The nature of the detected absorption}  
\label{sec:discussBLA}
The QSO sightline probing A1926 reveals three \lya\ components, all of which have line widths that indicate gas temperatures that could be as high as $10^{5-6}$ K.  This sightline lies on the outskirts of a cluster merger system, projected at 2 R$_{200}$ from A1926N, and shows an intriguingly symmetric arrangement of the \lya\ components (one component at $v = -937$ km s$^{-1}$ and another at $v = 1061$ km s$^{-1}$).  Furthermore, no absorption is detected between these two velocity extremes.  This is consistent with the findings of \citet{Yoon:2012yu}, who report a deficiency of \lya\ absorbers within a virial radius of Virgo and at similar velocities; however, the very-low redshift of Virgo hampers examining the absorption properties blueshifted relative to the cluster to look for such symmetry.  

In their hydrodynamical galaxy cluster simulations, \citet{Emerick:2015aa} find a bimodality in the velocities of \lya\ absorbers centred on the rest-frame velocities of their simulated clusters.  This effect arises because a significant fraction of their absorbers trace infalling filamentary material.  The BLAs we detect near A1926 are quite consistent with this behavior, suggesting that the absorbers may trace material accreting onto the cluster from the intergalactic medium and is either preheated to the warm-hot phase within the filament itself or at the accretion shock of the cluster environment \citep[$2-3$ \rvir;][]{Molnar:2009fk}.  It is notable that these BLA components have no associated \osix\ absorption and, indeed, infalling material from the IGM should have relatively low metallicity \citep[e.g.,][]{Hafen:2016aa,Fumagalli:2011qy}.  Unfortunately, the temperatures implied by the BLAs ($T\lesssim 10^{5.4}$ K) lie in a regime where non-equilibrium effects set in, which are in turn highly dependent on the gas metallicity \citet{Gnat:2007fk}.  Therefore, we are unable to constrain the metallicities of these absorbers.

The A1095 complex presents an interesting configuration where the X-ray data reveal the system to be an early-stage merger of two subclusters.  Furthermore, the QSO sightline pierces the interface region directly between the merging subclusters.  In contrast to the absorbers detected near A1926, we detect single narrow \hone\ component with $b = 16$ km s$^{-1}$, implying $T=10^{4.2}$ K under purely thermal broadening.  The curious presence of cool gas in this region that also shines in X-rays may be a result of the merger process.  For example, passing shocks can act to compress as well as strip gas, as seen in examples such as ESO 137-001 \citep{Sun:2007aa,Fumagalli:2014aa}, which displays prominent tails of star formation extending downstream as the galaxy passes through the ICM of the Norma cluster.  This phenomenon has been reproduced in hydrodynamic simulations \citep{Tonnesen:2012aa}, where substantial gas density enhancements occur in the wake of stripping events and then cool.  \citet{Tonnesen:2012aa} find that the level of star formation occurring within these dense pockets is sensitive to the pressure in the ICM, while \citet{Roediger:2014aa} showed that shocks propagating through the ICM due to a merger with another cluster can induce both thermal and ram pressure/density enhancements to levels sufficient to induce star formation.  The absorber we detect in this cluster merger interface region, which contains no galaxy in our survey data with $\rho< 300$ kpc and $\delta v < 400$ km s$^{-1}$ and thus does not appear in Figure \ref{fig:HIprofile}, has log N(\hone) = 14.93 $\cmt$, clearly well below what we might expect for a region where star formation might occur.  Possibilities for the nature of this absorber include the following: (1) The QSO sightline is probing a density enhancement produced by a merger shockwave out of detritus from a stripping event or, alternatively, from warm-hot gas located in the outer gas regions of a cluster.  (2) The absorption arises from cool gas that has been stripped from a galaxy near the merger interface and has remained cool at large distance from its source galaxy because of a large boost in the gas-galaxy relative velocity induced by the relative velocities of the merging clusters.   We note that a preliminary analysis of the galaxy spectra suggests that star formation activity is indeed enhanced along this merger interface, reminiscent of the ``Sausage" and ``Toothbrush" clusters \citep{Stroe:2015aa}, but we reserve a full analysis of those data for a subsequent paper.

\section{Summary and Conclusion}
\label{sec:summary}
We have presented HST/COS and optical spectroscopic observations to accompany our previously published X-ray imaging/spectroscopy of a sample of galaxy clusters selected for having background QSOs probing their ICM.  We leverage this combined dataset to study the CGM of the cluster galaxies as well as the warm and warm-hot phases of the ICM not detected in X-rays.  Our main results are summarized as follows:
\begin{enumerate}
\item The covering fraction, \fc, of \hone\ in the CGM of our cluster galaxies is $25^{+25}_{-15}\%$, much lower than the near-unity CGM \fc\ measured for field galaxies.  Furthermore, \fc\ for the for the cluster galaxies is significantly lower than that observed for group galaxies, which has in turn been measured to be lower than the field value.
\item In total, we detect three broad \lya\ (BLA) components at $\rho >$ \rtwo\ on the outskirts of one cluster (A1926) and two narrower components at $\rho <$ \rtwo\ of two clusters (A1095 and A2246).  The region of A1095 probed by our QSO sightline appears to be the interface of a subcluster merger.  
\item We detect no unambiguous \osix\ absorption in any our sightlines within $\Delta v = 2000$ km s$^{-1}$ of the cluster redshifts.
\item For the BLA components detected on the outskirts of A1926, we compare the mass of warm-hot $10^{5-6}$~K gas traced by BLA with the mass of X-ray traced gas.  We estimate that the warm-hot gas mass $\sim3\%$ of the hot gas mass, although much more could remain hidden if $T>10^{5.5}$ K.  Thus, at cluster-centric radii beyond \rtwo, such gas may comprise a significant contribution to the baryon content.
\item To enable follow-up studies of the clusters, we also publish our optical galaxy redshift surveys in these QSO fields in the Appendix. As the QSO spectra are also now public, these galaxy data should greatly increase the legacy value of the HST/COS data.
\end{enumerate}
Our study represents a burgeoning body of work using QSOs as absorption line probes of the densest regions in the low-redshift Universe, largely enabled by the sensitivity of the Cosmic Origins Spectrograph on the \emph{Hubble Space Telescope}.  Larger sample sizes are certainly required to place these intriguing findings of our work and other authors on firm statistical footing.  In particular, more sightlines systematically probing cluster-centric impact parameters from within to beyond \rtwo\ could illuminate the ionization and chemical states of the ICM as well as the transformation of the constituent galaxies' CGM relative to less dense regions of the Universe.

\section{Acknowledgements}  The authors would like to thank Mary Putman and Joo Heon Yoon for insightful discussion and sharing their data. Also, Benjamin Weiner provided extremely helpful advice for the Hectospec target selection strategy.  Hsiao-Wen Chen also provided helpful comments on an earlier draft.  The observations in this paper were obtained for HST programs 7778, 13833, 13342, and 13491, with financial support through NASA grants HST-GO-13342.002-A and HST-GO-13491.001-A from the Space Telescope Science Institute, which is operated by the Association of Universities for Research in Astronomy, Incorporated, under NASA contract NAS5-26555.  This work is partly based on observations taken at the MMT Observatory, a joint facility of the University of Arizona and the Smithsonian Institution.

\bibliography{Astronomy}

\section{Appendix: Spectroscopic galaxy sample}
We present the coordinates and redshifts for all galaxies spectroscopically measured, either through our own observations or those provided by the SDSS, in the QSO/cluster fields featured in this work.  In the table provided, the large majority of objects are drawn from our survey with MMT/Hectospec, for which the redshift measurement software provides quality scores that reflect the confidence of the measurement (see Section \ref{sec:galSpec}).  Although we use only objects with quality scores $\geq 3$ in our analysis, we include all objects for which we obtained spectra, which will be of interest to researchers who may wish to survey these fields further.  Also for these objects measured by Hectospec, we include the object classification provided by the measurement software.
\clearpage
\input{GalaxyData_edited.tex}

\end{document}

%% file: ClusterTable.tex
\begin{table*}
\caption{Selected properties of galaxy clusters and their QSO probes} 
\centering 
\begin{threeparttable} 
\begin{tabular}{ccccccc} 
\hline \hline 
Cluster &  $z_{\rm c}$\tnote{a} & log M$_{200}$ & $r_{200}$ & $\rho_{cl}$\tnote{b} & $\sigma_{cl}$\tnote{c} & QSO \\ 
 \ & \ & [log M$_{\odot}$] & [kpc] & [kpc] & [km s$^{-1}$] & \ \\ 
\hline 
Abell 1095W & 0.213 & 14.4 & 1240 & 865 & 860&SDSS J104741.75+151332.2 \\ 
Abell 1095E & 0.210 & 14.4 & 1190 & 1046 & 730& \\ 
Abell 1926S & 0.136 & 14.1 & 990 & 2411 & 400&2MASS J1431258+244220 \\ 
Abell 1926N & 0.136 & 14.0 & 940 & 1921 & 860& \\ 
Abell 2246 & 0.229 & 14.5 & 1300 & 491 & 780&HS1700+6416 \\ 
MaxBCG J217.84740+24.68382\tnote{d} & 0.097 & 13.9 & 890 & 156 & 520&2MASS J1431258+244220 \\ 
GMBCG J255.34805+64.23661 & 0.452 & 14.7 & 1400 & 1132 & 1050&HS1700+6416 \\ 
\hline 
\end{tabular} 
\begin{tablenotes}\item[a] Adopted cluster redshifts based on spectroscopic redshifts of BCGs.\item[b] Impact parameter defined as projected distance between QSO sightline and X-ray centroid assuming Hubble flow distance at redshift of BCG.\item[c] Velocity dispersions listed are approximations based on galaxies with spectroscopic redshifts projected within $r_{200}$ of each cluster; see Section \ref{sec:galaxies} text.\item[d] Values for $r_{200}$ and M$_{200}$ extracted from optical catalog \citep{Rykoff:2014aa}\end{tablenotes}\end{threeparttable} 
\label{tab:clusters} 
\end{table*}

%% file: QSO_Obs_edit.tex
\begin{table*}
\caption{Summary of QSO observations}
\begin{tabular}{ccccccr}
\hline
\hline
QSO & RA (J2000) & Dec (J2000) & $z_{\rm QSO}$ & T$_{\rm exp}$ & Instrument/Grating & Dataset IDs \\
\hline
SDSS J104741.75+151332.2 & 10 47 41.751 & +15 13 32.30 & 0.3858 & 24881.2 & COS G130M & LC8901010,LC8901020,LCKU09010 \\
~  & ~ & ~ & ~ & 18819.4 & COS G160M & LCKU10010,LCKU11010 \\
2MASS J1431258+244220 & 14 31 25.880 & +24 42 20.68 & 0.4069 & 14523.9 & COS G130M &
\multicolumn{1}{r}{\begin{tabular}[t]{@{}r@{}}LC8903010,LC8903020\\ LBS314010,LBS314020 \end{tabular}} \\
HS1700+6416 & 17 01 00.620 & +64 12 09.04 & 2.7407 & 175443.7 & COS G130M & LC8301010-LC8312010 \\
~ & ~ & ~ & ~ & 101968.0 & STIS E140M & O4SI02010-O4SI050A0 \\
\hline
\end{tabular}
\label{tab:QsoObs}
\end{table*}

%% file: LineMeasurementTable_VP.tex
\begin{table*}
\centering 
\begin{threeparttable} 
\caption{Voigt profile fitting results for absorbers within $\pm 2000$ km s$^{-1}$ of BCG redshifts and limits on potential absorbers within $\pm 600$ km s$^{-1}$ of cluster galaxy redshifts} 
\begin{tabular}{cccccc} 
\hline \hline 
Cluster &  $z_{\rm c}$ & log N(\hone) &  $b$ & $v$ & Detected lines \\ 
 \ &  \ & [cm$^{-2}$] & [km/s] & [km/s] & \ \\ 
\hline 
A1095 & 0.2108 & 14.93 $\pm$ 0.03 & 16 $\pm$ 1 & 36 $\pm$ 1 & Ly$\alpha$,Ly$\beta$,Ly$\gamma$,Ly$\delta$,Ly$\epsilon$ \\ 
A1926 & 0.1358 & 13.66 $\pm$ 0.04 & 69 $\pm$ 9 & -937 $\pm$ 6 & Ly$\alpha$,Ly$\beta$ \\ 
A1926 & 0.1358 & 13.12 $\pm$ 0.12 & 61 $\pm$ 24 & -736 $\pm$ 15 & Ly$\alpha$,Ly$\beta$ \\ 
A1926 & 0.1358 & 13.18 $\pm$ 0.11 & 71 $\pm$ 23 & 1061 $\pm$ 16 & Ly$\alpha$,Ly$\beta$ \\ 
A2246 & 0.2289 & 13.97 $\pm$ 0.02 & 25 $\pm$ 1 & 260 $\pm$ 1 & Ly$\alpha$,Ly$\beta$,Ly$\gamma$ \\ 
A2246 & 0.2289 & $<$13.30\tnote{1} & 32 & -197 & - \\ 
A2246 & 0.2289 & $<$13.18\tnote{1} & 22 & -116 & - \\ 
A2246 & 0.2289 & $<$13.50\tnote{1} & 11 & -57 & - \\ 
A2246 & 0.2289 & 14.63 $\pm$ 0.01 & 24 $\pm$ 1 & -1612 $\pm$ 1 & Ly$\alpha$,Ly$\beta$,Ly$\gamma$,Ly$\delta$ \\ 
GMBCG J255.55+64.23 & 0.4523 & $<$14.08\tnote{1} & 20 & 681 & - \\ 
GMBCG J255.55+64.23 & 0.4523 & $<$13.78\tnote{1} & 30 & 761 & - \\ 
GMBCG J255.55+64.23 & 0.4523 & $<$13.50\tnote{1} & 30 & 856 & - \\ 
GMBCG J255.55+64.23 & 0.4523 & 14.72 $\pm$ 0.01 & 18 $\pm$ 1 & -1856 $\pm$ 1 & Ly$\alpha$,Ly$\beta$,Ly$\gamma$,Ly$\delta$,Ly$\epsilon$ \\ 
\hline 
\label{tab:ICMmeas} 
\end{tabular} 
\begin{tablenotes} 
\item[1] Upper limit on column density derived by fitting Voigt profiles to potentially blended absorption.  See Section \ref{sec:resultsCGM} and Figures \ref{fig:Stackplot3} and \ref{fig:Stackplot4}.  
\end{tablenotes} 
\end{threeparttable} 
\end{table*}

%% file: LineMeasurementTable.tex
\begin{table*}
\caption{Column densities of \hone\ and \osix\ associated with individual cluster galaxies} 
\centering 
\begin{threeparttable} 
\begin{tabular}{ccccc} 
\hline \hline 
Cluster &  $z_{\rm gal}$ & $\rho$ & log N(\hone) &  log N(\osix) \\ 
 \ & \ & [kpc] & [cm$^{-2}$] \\ 
\hline 
MaxBCG J217.55+24.68 & 0.0939 & 269 & $<$12.97 & - \\ 
MaxBCG J217.55+24.68 & 0.0960 & 129 & $<$12.90 & - \\ 
MaxBCG J217.55+24.68 & 0.0993 & 193 & $<$12.91 & - \\ 
MaxBCG J217.55+24.68 & 0.0964 & 117 & $<$12.91 & - \\ 
MaxBCG J217.55+24.68 & 0.0961 & 236 & $<$12.92 & - \\ 
MaxBCG J217.55+24.68 & 0.0960 & 131 & $<$12.92 & - \\ 
MaxBCG J217.55+24.68 & 0.0963 & 210 & 12.97 $\pm$ 0.13 & - \\ 
A1095 & 0.2130 & 234 & $<$13.53 & $<$13.11 \\ 
A1095 & 0.2135 & 245 & $<$13.54 & $<$13.12 \\ 
A2246 & 0.2285 & 44 & $<$14.20\tnote{1}  & $<$13.80 \\ 
GMBCG J255.55+64.23 & 0.4560 & 73 & $<$14.33 & $<$14.62 \\ 
\hline 
\end{tabular} 
\begin{tablenotes}\item[1] Upper limit includes the column density of unambiguously detected \hone\ and any possible contribution of \hone\ to nearby blended absorption features\end{tablenotes}\end{threeparttable}\label{tab:CGMmeas} 
\end{table*}

%% file: GalaxyData_edited.tex
\begin{table}
\begin{minipage}{\linewidth} \centering Table A1: Galaxy survey data in QSO/cluster fields. \\ 
\begin{tabular}{ccccccccc}
\hline
Field & Galaxy Name & RA (J2000) & Dec (J2000) & z & $\sigma_z$ & Quality & Object Class & Source \\
 &  & (deg) & (deg) &  &  &  &  &  \\
\hline
A1095 & J104645.19+153230.52 & 161.688293 & 15.541811 & 0.1457 & 0.0005 & 4 & GALAXY & Hectospec \\
A1095 & J104835.24+151438.78 & 162.146820 & 15.244105 & 0.3237 & 0.0005 & 4 & GALAXY & Hectospec \\
A1095 & J104751.4+151658.04 & 161.964188 & 15.282789 & 0.2114 & 0.0005 & 4 & GALAXY & Hectospec \\
A1095 & J104806.03+151851.01 & 162.025116 & 15.314170 & 0.2082 & 0.0005 & 4 & GALAXY & Hectospec \\
A1095 & J104832.31+151223.49 & 162.134644 & 15.206525 & 0.1785 & 0.0005 & 4 & GALAXY & Hectospec \\
A1095 & J104835.46+151925.10 & 162.147781 & 15.323639 & 0.2478 & 0.0005 & 4 & QSO & Hectospec \\
A1095 & J104835.14+152359.25 & 162.146423 & 15.399792 & 0.0855 & 0.0005 & 4 & GALAXY & Hectospec \\
A1095 & J104818.04+151740.33 & 162.075180 & 15.294536 & 0.2124 & 0.0005 & 4 & GALAXY & Hectospec \\
A1095 & J104811.34+151917.88 & 162.047241 & 15.321633 & 0.2086 & 0.0005 & 4 & GALAXY & Hectospec \\
A1095 & J104802.8+151636.66 & 162.011642 & 15.276850 & 0.2155 & 0.0005 & 4 & GALAXY & Hectospec \\
A1095 & J104755.77+151558.68 & 161.982391 & 15.266300 & 0.2147 & 0.0005 & 4 & GALAXY & Hectospec \\
A1095 & J104838.57+152157.11 & 162.160706 & 15.365864 & 0.3664 & 0.0005 & 4 & GALAXY & Hectospec \\
A1095 & J104759.63+151530.77 & 161.998474 & 15.258547 & 0.2123 & 0.0005 & 4 & GALAXY & Hectospec \\
A1095 & J104841.98+151818.68 & 162.174911 & 15.305189 & 0.1562 & 0.0005 & 4 & GALAXY & Hectospec \\
A1095 & J104810.4+151429.43 & 162.043320 & 15.241508 & 0.2174 & 0.0005 & 4 & GALAXY & Hectospec \\
A1095 & J104829.44+151701.14 & 162.122681 & 15.283650 & 0.3869 & 0.0005 & 4 & GALAXY & Hectospec \\
A1095 & J104803.22+151433.37 & 162.013428 & 15.242602 & 0.2125 & 0.0005 & 4 & GALAXY & Hectospec \\
A1095 & J104749.01+151306.74 & 161.954224 & 15.218539 & 0.2113 & 0.0005 & 4 & GALAXY & Hectospec \\
A1095 & J104847.07+151654.91 & 162.196106 & 15.281919 & 0.1547 & 0.0005 & 4 & GALAXY & Hectospec \\
A1095 & J104912.22+151907.91 & 162.300903 & 15.318864 & 0.4238 & 0.0005 & 4 & GALAXY & Hectospec \\
A1095 & J104758.96+151507.34 & 161.995667 & 15.252039 & 0.2097 & 0.0005 & 4 & GALAXY & Hectospec \\
A1095 & J104816.24+151247.78 & 162.067657 & 15.213272 & 0.1325 & 0.0005 & 4 & GALAXY & Hectospec \\
A1095 & J104915.41+151551.26 & 162.314178 & 15.264239 & 0.4386 & 0.0005 & 4 & GALAXY & Hectospec \\
A1095 & J104823.14+151123.38 & 162.096405 & 15.189828 & 0.3044 & 0.0005 & 4 & GALAXY & Hectospec \\
A1095 & J104920.4+151536.55 & 162.335022 & 15.260153 & 0.1691 & 0.0005 & 4 & GALAXY & Hectospec \\
A1095 & J104834.68+152425.54 & 162.144516 & 15.407094 & 0.0839 & 0.0005 & 4 & GALAXY & Hectospec \\
A1095 & J104829.86+152623.13 & 162.124420 & 15.439758 & 0.1766 & 0.0005 & 4 & GALAXY & Hectospec \\
A1095 & J104857.71+152846.49 & 162.240479 & 15.479581 & 0.1680 & 0.0005 & 4 & GALAXY & Hectospec \\
A1095 & J104804.35+153416.81 & 162.018097 & 15.571336 & 0.2117 & 0.0005 & 4 & GALAXY & Hectospec \\
A1095 & J104812.13+145607.38 & 162.050537 & 14.935384 & 0.4357 & 0.0005 & 4 & GALAXY & Hectospec \\
A1095 & J104746.57+152258.17 & 161.944061 & 15.382825 & 0.1910 & 0.0005 & 4 & GALAXY & Hectospec \\
A1095 & J104741.25+152056.76 & 161.921875 & 15.349100 & 0.1550 & 0.0005 & 4 & QSO & Hectospec \\
A1095 & J104745.14+152127.40 & 161.938065 & 15.357611 & 0.3000 & 0.0005 & 4 & GALAXY & Hectospec \\
A1095 & J104742.06+153443.42 & 161.925232 & 15.578728 & 0.3582 & 0.0005 & 4 & GALAXY & Hectospec \\
A1095 & J104817.84+153523.94 & 162.074326 & 15.589983 & 0.1778 & 0.0005 & 4 & GALAXY & Hectospec \\
A1095 & J104755.95+152834.81 & 161.983124 & 15.476336 & 0.2095 & 0.0005 & 4 & GALAXY & Hectospec \\
A1095 & J104806.14+153824.32 & 162.025574 & 15.640089 & 0.2107 & 0.0005 & 4 & GALAXY & Hectospec \\
A1095 & J104801.52+153252.46 & 162.006348 & 15.547906 & 0.2141 & 0.0005 & 4 & GALAXY & Hectospec \\
A1095 & J104803.8+145941.52 & 162.015823 & 14.994866 & 0.1606 & 0.0005 & 4 & GALAXY & Hectospec \\
A1095 & J104812.61+150818.43 & 162.052536 & 15.138453 & 0.2129 & 0.0005 & 4 & GALAXY & Hectospec \\
A1095 & J104812.48+152237.13 & 162.052017 & 15.376981 & 0.2141 & 0.0005 & 4 & GALAXY & Hectospec \\
A1095 & J104833.71+153205.94 & 162.140472 & 15.534984 & 0.1782 & 0.0005 & 4 & GALAXY & Hectospec \\
A1095 & J104749.8+152007.45 & 161.957504 & 15.335402 & 0.2130 & 0.0005 & 4 & GALAXY & Hectospec \\
A1095 & J104819.45+152746.00 & 162.081055 & 15.462778 & 0.1781 & 0.0005 & 4 & GALAXY & Hectospec \\
A1095 & J104748.25+151832.05 & 161.951019 & 15.308903 & 0.2100 & 0.0005 & 4 & GALAXY & Hectospec \\
A1095 & J104747.53+151630.04 & 161.948059 & 15.275011 & 0.2965 & 0.0005 & 4 & GALAXY & Hectospec \\
A1095 & J104806.6+152419.16 & 162.027512 & 15.405322 & 0.1736 & 0.0005 & 4 & GALAXY & Hectospec \\
A1095 & J104815.25+145059.97 & 162.063538 & 14.849992 & 0.2675 & 0.0005 & 4 & GALAXY & Hectospec \\
\hline
\end{tabular}
\end{minipage}
A full electronic version is available from the journal website.
\end{table}